\documentclass{JHEP3}
\usepackage{epsfig}
\usepackage{amsmath}
\usepackage{mathrsfs}

\newcommand{\be}{\begin{equation}}
\newcommand{\ee}{\end{equation}}
\newcommand{\ba}{\begin{eqnarray}}
\newcommand{\ea}{\end{eqnarray}}
\newcommand{\ra}{\rangle}
\newcommand{\la}{\langle}
\newcommand{\st}{\scriptstyle}
\newcommand{\sst}{\scriptscriptstyle}
\newcommand{\nn}{\nonumber\\}

\title{Group theoretical approach to \\ quantum fields in de Sitter space \\ 
II. The complementary and discrete series}

\author{Euihun Joung\\
	APC, Universit\'e Paris VII, B\^at. Condorcet,  
	75205 Paris Cedex 13, France\\
	joung@apc.univ-paris7.fr}
\author{Jihad Mourad\\
	APC, Universit\'e Paris VII, B\^at. Condorcet,
	75205 Paris Cedex 13, France\\
	mourad@apc.univ-paris7.fr}
\author{Renaud Parentani\\
	LPT, Universit\'e Paris XI, B\^at. 210,
	91405 Orsay Cedex, France\\
	parenta@th.u-psud.fr}

\abstract{
We use an  algebraic approach based on representations of  de
Sitter group to
construct covariant quantum fields in arbitrary dimensions. 
We study the complementary and the 
discrete series which correspond to light and massless
fields and which lead new feature with respect to 
the massive principal series
we previously studied (\hepth{0606119}). 
When considering the complementary series,
we make use of a non-trivial scalar product in order to 
get  local expressions in the position representation.
 Based on these, we construct a family of covariant canonical fields
 parametrized by SU$(1,1)/$U$(1)$.
Each of these correspond to the dS invariant alpha-vacua.
The behavior of the modes at asymptotic times brings another
difficulty as it is incompatible with the usual definition of 
the \emph{in} and \emph{out} vacua.
 We propose a generalized notion of these vacua which reduces to
the usual conformal vacuum in the conformally massless limit.
 When considering the massless discrete series we find
that no covariant  field obeys the canonical commutation relations.
To further analyze this singular case, we consider the massless limit of 
the complementary scalar fields we previously found. 
We obtain canonical fields with a deformed representation by zero modes.
 The zero modes have a dS invariant vacuum with singular norm.
 We propose a regularization by a compactification of the scalar field
 and a dS invariant definition of the vertex operators. The resulting
 two-point functions are dS invariant and have a universal logarithmic infrared
 divergence. 
 }

\keywords{Space-Time Symmetries, Global Symmetries, dS vacua in string theory}

\begin{document}

\section{Introduction}

Since the works of Dirac and Wigner \cite{Wigner:1939cj}, 
particles in flat spacetime can be considered as unitary irreducible representations (UIR)
of the Poincar\'e group. Moreover, free quantum field operators $\Phi(x)$ can
be constructed in a unique way from these UIR \cite{Weinberg:1995mt,Weinberg:1996kw}. 
The key property in this construction is the covariance of the field operator:
\be
	\Phi(x')=U(\Lambda)\,\Phi(x)\,U^{\dagger}(\Lambda),\label{cov}
\ee
where we only consider scalar fields 
and $x'$ is the image of $x$ under
the Poincar\'e transformation $\Lambda$.

In a previous work \cite{Joung:2006gj}, hereafter cited as I,
we presented an algebraic construction of quantum field theories (QFT)
on the $n$-dimensional de Sitter space $dS_n$
based on the UIRs of the de Sitter isometry group SO$_{0}(1,n)$.
The UIR were first analyzed  by Bargmann \cite{Bargmann:1946me}
for $n=2$, Gelfand and Naimark for $n=3$ \cite{gel}, Thomas, Newton and
Dixmier for $n=4$ \cite{th,nt,rep2}. Generalization to all $n$ were studied in 
\cite{rep3,rep4,rep5,rep6}.
Starting with an UIR, we first construct the corresponding Fock space.
The vacuum is the trivial representation, 
the one-particle states are elements of the UIR,
and the $n$-particle states are the symmetrized tensor product of 
$n$ copies of the UIR.
The dS group has an induced representation $U$ on this Fock space.
We then define a free local field operator by a linear superposition
of creation and annihilation operators subject to the covariance property 
(\ref{cov}) with $\Lambda$ now being an element of SO$_0(1,n)$. 
The scalar representations are of 
three types characterized by the eigenvalue of the quadratic Casimir operator ${\cal C}$:
the principal series has ${\cal C}\le-(n-1)^2/4$, 
the complementary series has $-(n-1)^2/4\le{\cal C}<0$ and finally
the discrete series has  ${\cal C}=k(k+n-1)$ with $k\in{\mathbb Z}_{\ge0}\,$.
The eigenvalue of ${\cal C}$ can be interpreted physically as 
the minus mass squared : ${\cal C}=-M^2$ 
(or more generally with a curvature coupling term, ${\cal C}=-(M^2+\xi\,{\cal R})$).
In I, we studied the principal series and we showed that 
the construction gives rise to a family of canonical fields parametrized by
a SU$(1,1)/$U$(1)$  moduli space. In the usual field 
theoretical treatment, this moduli space corresponds to that of alpha-vacua \cite{Allen:1985ux}.
In our approach, the moduli space stems from 
a first order differential equation expressing the invariance
of the field operator under transformations leaving fixed a given point in dS space.
This first order differential equation turns out to be singular
(on the event horizons centered about that point) thereby 
giving rise to two independent complex solutions. 
Furthermore, we showed that the Klein-Gordon equation 
and the canonical commutation relations
are consequences of the covariance requirement.

The aim of the present work is
to extend this approach to the other two series.
Concerning the complementary series, the main
difference with respect to the principal one is 
that the realization of the representation with functions on the sphere
is no more unitary with respect to the standard ${\cal L}^2$ scalar product.
This realization was very convenient in I because
the generators are local differential operators
and the finite transformations easy to determine.
The unitarity of the complementary series can be recovered by
the use of a different scalar product which we first determine.
The construction then follows the same general line as before leading
once again to a family of free field
operators labeled by the moduli space SU$(1,1)/$U$(1)$.
An important physical difference with respect to the principal series
exists however, it concerns the 
 different behavior of the field operators in the asymptotic past and
future. This forbids the usual definition of the \emph{in} and \emph{out} Mottola-Schwinger vacua \cite{Mottola:1984ar}. 
We propose a new definition of asymptotic vacua
by appropriately factorizing the field operator and choosing the time coordinate.
Our construction generalizes the conformal \emph{in} and \emph{out} vacua valid
for the conformally coupled scalar fields with ${\cal C}=-n(n-2)/4\,$.

As of the discrete series, we consider only
the physically interesting case of the first 
discrete series with ${\cal C}=0$, i.e. the massless case.
Other discrete series correspond to the tachyonic fields 
which are physically less relevant (see \cite{Folacci:1996dv} for an example).
The massless scalar case is peculiar in the usual field theoretic approach
since the dS invariant two-point function diverges in the limit $M\to0$ \cite{Ford:1977in}.
This was interpreted as there is no dS invariant vacuum in the massless case \cite{Allen:1985ux}
and the vacuum states breaking dS group but
preserving its subgroup were considered in 
\cite{Allen:1985ux,Allen:1987tz,Polarski:1991ek,Kirsten:1993ug}.
 It was suggested that this divergence is related to the additional symmetry
 which the massless theory acquires : the symmetry under the constant addition 
on the field operator.
 In order to implement this symmetry appropriately,
 the BRST quantization in the Euclidean dS space \cite{Folacci:1992xc} and 
 the Gupta-Bleuler quantization \cite{Gazeau:1999mi}
 was studied and also the invariant observables under this symmetry
 rather than field operators were considered: 
 difference of fields in different spacetime points \cite{Kirsten:1993ug}.
In the approach of the present paper,
the construction for the massless discrete series leads to 
quite different results from that of complementary series.
The field operator so obtained turns out to be unique up to an overall complex constant
and more importantly, it does not satisfy the canonical commutation relations.
This QFT gives physically unacceptable effects:
its coupling to an Unruh detector \cite{Unruh:1975gz}  leads to
an infinite temperature.
We then explore the limit towards the massless field starting from the
canonical massive fields of the complementary series.
The limit gives rise a Fock space which is larger than that obtained
from the discrete UIR, the additional part being due to the zero modes.
The representation is also deformed by the zero modes 
in a way which we explicitly determine.
The zero modes however render the  vacuum not normalizable.
We then cure this problem by compactifying the scalar field
on a circle $\Phi=\Phi+2\pi L$. 
In order to get observables invariant under this rotation,
we consider a vertex operator which is a dS invariant
regularization of $\exp(i\Phi/L)$. We compute its
two-point function and show that it presents a 
universal logarithmic infrared divergence for all dimensions.
We thus recover in dS invariant way the results of  \cite{Ratra:1984yq}
where these divergences were shown to lead to
the restoration of broken symmetries.

The plan of the paper is as follows.
From Section 2 to Section 5, we study
the two dimensional case. 
In Section 6, we generalize the results to arbitrary dimensions.
In Section 2, we describe the representations and the associated scalar product.
We also examine the massless limit from the complementary series.
In Section 3, we construct the scalar field of the complementary series
and determine the Bunch-Davies vacuum \cite{BD} and our 
generalization of the \emph{in} and \emph{out} vacua. 
Section 4 is devoted to the massless case and Section 5
to the massless limit of the scalar field of the complementary series.
Several Appendices contain technical details used in the text.

\section{The SO$_0$(1,2) group}

We first concentrate on the two-dimensional de Sitter space $dS_2$
and the starting point of our approach is the UIR of the SO$_0(1,2)$
group, the group of linear transformations with determinant $1$
which leaves $-(X^0)^2+(X^1)^2+(X^3)^2$ invariant
and which is connected to the identity. This is the isometry group
of $dS_2$. The arbitrary dimensional case, SO$_0(1,n)$
will be treated at the end of the paper. 
Let $\cal J$ be the generator of
the rotation subgroup and ${\cal K}_1$ and ${\cal K}_2$ the two boosts. They
verify the commutation relations:
\be
   [\,{\cal J,\, K}_1\,]=i{\cal K}_2\,,\qquad [\,{\cal J,\,K}_2\,]=
   -i{\cal K}_1\,,\qquad [\,{\cal K}_1,\,{\cal K}_2\,]=-i{\cal J}\,.
\ee
The quadratic Casimir operator
\be
   {\cal C=J}^2-{\cal K}_1^2-{\cal K}_2^2\,,
\ee
commutes with all the
generators and is constant on irreducible representation.
Bargmann \cite{Bargmann:1946me} classified the UIR according to the value of $\cal C$ and the
eigenvalues $m$ of $\cal J$:
\begin{itemize}
	\label{class}
   \item
       (i) the principal series with ${\cal C}\le-{1\over4}$,\quad
       $m=0,\pm 1,\dots$ or $m=\pm{1\over 2},\pm{3\over 2},\dots$;
   \item
       (ii) the complementary series with $-{1\over 4}<{\cal C}<0$ and $m=0,\pm1,\dots$;
   \item
       (iii) the two discrete series $D^\pm_k$, with ${\cal C}=k(k+1)$ 
       where $k$ is a non-negative integer or half integer,
       and with $\pm m=k+1,k+2,\dots$ where the $+$ ($-$) 
       sign characterizing $D^+_k$ ($D^-_k$)
\end{itemize}

 In all these cases, writing
 ${\cal C}=s^2-{1/4}$,
 the three generators can be represented in 
the basis of the  eigenstates of ${\cal J}$ as
\be
   {\cal J}\,|m)=m\,|m)\,,\qquad
   {\cal K}_{\pm}\,|m)=\pm i\left\{\,m
   \pm\left(s+{1\over 2}\right)\right\}|m \pm 1)\,.\label{repre}
\ee
where the raising and lowering operators are defined as 
${\cal K}_{\pm}={\cal K}_1\pm i{\cal K}_2$. 
We have adopted these expressions because 
 the action of ${\cal K}_{\pm}$
is linear in $m$. Therefore, they give rise to
first order differential operators in the position representation,
 that is when acting on  functions on the circle.
Indeed, defining
\be
	|\phi)={1\over \sqrt{2\pi}}\sum_{m=-\infty}^{\infty}e^{-i m\phi}\,|m)\,,
\ee
the action of the generators on
an arbitrary ket $|\Psi)$ is given by
\ba
   (\phi\vert\, {\cal J}\,\Psi) &=&-i{d \over
   d\phi}(\phi\vert\Psi)\,
   ,\nonumber \\
   (\phi\vert\,  {\cal K}_1\,\Psi)
   &=&i\left\{ \sin\phi{d \over d\phi}+\left(s+{1\over 2}\right)\cos\phi\right\}(\phi\vert\Psi)
   =i\left[{1\over 2}\left\{\sin\phi,{d \over d\phi}\right\}-s\cos\phi\right](\phi\vert\Psi)\,,
   \nonumber\\
   (\phi\vert\,  {\cal K}_2\,\Psi)
   &=&i\left\{-\cos\phi{d \over d\phi}+\left(s+{1\over 2}\right)\sin\phi\right\}(\phi\vert\Psi)
   =-i\left[{1\over 2}\left\{\cos\phi,{d \over d\phi}\right\}+s\sin\phi\right](\phi\vert\Psi)\,.
   \nonumber\\
\ea
In the sequel, it will be also useful to have the action of finite
transformations. They are given by
\ba
   (\phi\vert \,e^{i\theta {\cal J}}\, \Psi)
   &=& (\phi+\theta\vert\Psi)\,,
   \nonumber\\
   (\phi\vert \,e^{i\rho {\cal K}_1}\,\Psi)
   &=&(\cosh\rho +\sinh\rho\cos\phi)^{-s-1/2} \,(\phi_1\vert\Psi)\,,
   \nonumber \label{te}\\
   (\phi\vert \,e^{i\lambda {\cal K}_2}\,\Psi)
   &=& (\cosh\lambda +\sinh\lambda\sin\phi)^{-s-1/2}\, (\phi_2\vert\Psi)\,.
\ea
where
\ba
   \cos\phi_1 &=& {\cos\phi  \cosh\rho+\sinh\rho\over \cosh\rho+\sinh\rho \cos\phi}\,,\qquad
   \sin\phi_1={\sin\phi \over \cosh\rho +\sinh\rho\ \cos\phi}\,,
   \nonumber\\
   \cos\phi_2&=&{\cos\phi
   \over \cosh\lambda +\sinh\lambda\ \sin\phi}\,,\qquad \sin\phi_2={\sin\phi \cosh\lambda+\sinh\lambda
   \over \cosh\lambda +\sinh\lambda\ \sin\phi}\,.
\ea
The choice adopted in eq.(\ref{repre})  yields the above 
simple expressions for the finite transformations 
and will be easily generalized to higher dimensions.
However, it should be noticed that
for the discrete and
the complementary representations,
the generators are not Hermitian with 
respect to the standard ${\cal L}^2$ scalar product,
we denote $(\cdot|\cdot)$,
for the discrete and
the complementary representations.
In fact,  denoting  ${\cal U}^{(s)}$ 
the representation (\ref{repre}) for some $s$
of three generators of SO$_0(1,2)$, we have
\be
      (\Psi\vert \ {\cal U}^{(s)}\,\Psi')=(\,{\cal U}^{(-s^*)}\,\Psi\vert\Psi')\,,
      \label{sp}
\ee
where ${\cal U}^{(s)}$ 
designates any of three generators of SO$_0(1,2)$
in the representation (\ref{repre}).
Hence  the ${\cal U}^{(s)}$ are Hermitian with respect to the ${\cal L}^2$ scalar
product  only if $s$ is purely imaginary, e.g. for
the principal series.
For the other
two series we should define a new scalar product
with respect to which the generators are Hermitian.

\subsection{The complementary series}

The complementary series is obtained by taking $s$ real and 
belonging to the interval 
$-{1/2}<s<{1/2}\,$.
It turns out that 
$s$ and $-s$ are equivalent representations because
there exists an intertwiner  $Q$ such that
 \be
      Q\ {\cal U}^{(s)}\,=\,{\cal U}^{(-s)}\,Q\,.
      \label{int}
\ee
This intertwiner defines a new scalar product, 
denoted $\la\cdot|\cdot\ra$, by 
\be
	\la\Psi\,|\,\Psi'\ra\equiv(\Psi\,\vert\,Q\,\Psi')\,,
\ee
and with respect to which, the generators are Hermitian:
\ba
     \la\Psi\,|\ {\cal U}^{(s)}\,\Psi'\ra&=&(\Psi\,\vert\,Q\ {\cal U}^{(s)}\,\Psi')
     \,=\,(\Psi\,\vert\ {\cal U}^{(-s)}\,Q\,\Psi')
     \nonumber\\
     &=&(\,{\cal U}^{(s)}\,\Psi\,\vert\,Q\,\Psi')
     \,=\,\la\,{\cal U}^{(s)}\,\Psi\,|\,\Psi'\ra\,.
     \label{Hp}
\ea
From the the action of the generators and the
definition of $Q$ we deduce the scalar product of states 
$|m)$ and $|n)$, i.e. the matrix element of $Q$,
 as
\be
	\la m|n\ra\,=\,(m\vert Q\,n)\,=\,
	\la m_0|m_0\ra\,
	\frac{\Gamma\left(m_0+s+\frac12\right)}{\Gamma\left(m_0-s+\frac12\right)}\,
	\frac{\Gamma\left(m-s+\frac12\right)}{\Gamma\left(m+s+\frac12\right)}\,
	\delta_{m,n}\,\equiv Q_m\delta_{m,n}\, .
	\label{Q_comp}
\ee
For $-1/2<s<1/2$, this scalar product is
 positive and regular 
for all states $|m)\,$. For $0<s<1/2$ 
the scalar product of two position eigenvectors can be obtained by 
Fourier transformation. One finds \cite{Bargmann:1946me} 
\ba
      	Q(\phi-\phi')&\equiv& (\phi\vert Q\,\phi')
	=\frac1{2\pi}\sum_{m=-\infty}^{\infty}e^{im(\phi-\phi')}\,Q_m
	\nonumber\\
	&=&\frac{\la m_0|m_0\ra}{\sqrt{4\pi}}\,
	\frac{\Gamma\left(m_0+s+\frac12\right)}{\Gamma\left(m_0-s+\frac12\right)}\,
	\frac{\Gamma\!\left(-s+\frac12\right)}{\Gamma(s)}
      	\,\Big(\sin^2{(\phi-\phi')\over 2}\Big)^{s-\frac12}\,.
	\label{sp comp}
\ea
This can be also obtained from the following differential equation:
\be
	(\phi|\left({\cal K}_+^{(-s)}\,Q-Q\,{\cal K}_+^{(s)}\right)\phi')=
	\left\{\sin\frac \varphi2\,\frac d{d\varphi}-
	\left(s-\frac12\right)\cos\frac \varphi2\right\}Q(\varphi)=0\,,
	\qquad \varphi=\phi-\phi'\,.
\ee
In the region $-1/2<s<0$, the above expression is singular
as can be seen from the behavior of the last factor in the coincident point limit. 
To our knowledge this case has not been studied in the literature. 
To obtain an expression valid for all values of $s$ in the complementary series 
we should consider the above kernel $Q(\phi-\phi')$ as a  distribution.
This distribution is given by
\ba
	Q(\phi-\phi')&=&\la m_0|m_0\ra\,
	\frac{\Gamma\left(m_0+s+\frac12\right)}{\Gamma\left(m_0-s+\frac12\right)}\,
	\frac{\Gamma\!\left(-s+\frac12\right)}{(4\pi)^\frac12\,\,\Gamma(s)}\times
	\nonumber\\
	&&\times
	\lim_{\epsilon\to 0^+}
	\left[\Big(\sin^2{(\phi-\phi')\over 2}+{\epsilon\over 2s}\Big)
	\Big(\sin^2{(\phi-\phi')\over 2}+{\epsilon}\Big)^{s-{3\over 2}}\right].
	\label{ps}
\ea
 This expression is now valid for $-1/2<s<1/2$
and reduces to eq.(\ref{sp comp}) for $0<s<1/2$.
Furthermore, when $s\to0$ it reduces, as it should,  to
\be
	Q(\phi-\phi')=\la m_0|m_0\ra\,\delta(\phi-\phi')\,,
\ee
which is the ${\cal L}^2$ kernel of the principal series.
As we shall see below, this generalization of $Q$ to negative values of $s$
will be useful to study
the massless case as a limiting procedure from the complementary series.

\subsection{The massless case}

The massless representations have a vanishing Casimir,
i.e. $s^2 = 1/4$, and constitute the first discrete representations 
$D_{k=0}^\pm$ in \textref{class}{the classification of UIR}.
The positive series $D_0^+$ has only states with positive $m$ 
whereas the negative series $D_0^-$ with only negative $m$. 
Any state $|\Psi)$ in each series $D_0^\pm$ can 
thus be expanded as
\be
	|\Psi)=\sum_{m=1}^{\infty}c_{\pm m}\,|\pm m)\,.
\ee
As can be seen from eq.(\ref{repre})
the expressions  greatly simplify for $s=-1/2$:
\be
   	{\cal J}\,|m)=m\,|m)\,,
	\qquad
   	{\cal K}_{\pm}\,|m)=\pm i\,m\,|m \pm 1)\,.
\ee
They are Hermitian provided that the scalar product for $D_0^\pm$ is given by
\be
	\langle n|m\ra=\frac{\langle m_\pm|m_\pm\rangle}{m_\pm}\,m\,\delta_{m,n}\,,
\ee
where $m_+$(resp. $m_-$) is an arbitrary positive(resp. negative) integer.
It should be noticed  that the zero mode $|m=0)\equiv i\,{\cal K}_\mp|\pm1)$ 
has vanishing norm and so does not belong to the UIRs
of $D^\pm_0$.

To be able to use the position representation
of the generators acting on $D_0^\pm$, we consider the (larger)
Hilbert space formed by the
${\cal L}^2$ functions on the circle. This space  includes all modes $m$
 and allows the  following decomposition of any element of 
$D^\pm_0$  as
\be
	\Psi(\phi)\equiv(\phi|\Psi)=\sum_{m=1}^{\infty}e^{\pm im\phi}\,c_{\pm m}\,.
	\label{phys}
\ee
In the position basis, the scalar product between two elements of $D_0^\pm$
is 
\be
 	\la \Psi|\Psi'\ra=-i\,
	{\langle m_\pm|m_\pm\rangle\over m_\pm}
	\int_0^{2\pi}{d\phi}\,(\Psi|\phi)\,\frac{d}{d\phi}(\phi|\Psi')\,.
 \ee
In this enlarged Hilbert space, the action
of the three generators is given by
\be
   	(\phi\vert\,{\cal J}\,\Psi)
	=-i{d \over d\phi}(\phi\vert\Psi)\,,
	\quad
   	(\phi\vert\,{\cal K}_1\,\Psi)
   	=i\sin\phi{d \over d\phi}(\phi\vert\Psi)\,,
	\quad
	(\phi\vert\,{\cal K}_2\,\Psi)
   	=-i\cos\phi{d \over d\phi}(\phi\vert\Psi)\,.
\ee

\subsection{The massless limit}

The  limit $s\rightarrow {-1/2}$ of the complementary series
leads at a first sight to three irreducible representations: 
$D^{+}_0$ with $m\ge1$, $D^-_0\,$, with $m\le-1$ and the trivial representation with $m=0\,$.
In fact, writing $s=-1/2+\varepsilon$, 
the action of ${\cal K}_\mp$ on the normalized 
states with $m=\pm 1$ is
\be
	{\cal K}_\mp \frac{|\pm1)}{\sqrt{\la\pm 1|\pm1\ra}}=
	-i\sqrt{\varepsilon}\,\frac{|0)}{\sqrt{\la0|0\ra}}\,,
	\qquad
	{\cal K}_\pm \frac{|0)}{\sqrt{\la0|0\ra}}=\pm
	i\sqrt{\varepsilon}\,\frac{|\pm1)}{\sqrt{\la\pm1|\pm1\ra}}\,.
\ee
These equations shows that 
the representation splits indeed
into the three irreducible
representations mentioned above. 
The scalar product however has a singular limit
since eq.(\ref{Q_comp}) gives
\be
	\la 0|0\ra
	=\frac{\la m_0|m_0\ra}{m_0}\,\varepsilon+{\cal O}(\varepsilon^2)\,.
	\label{0limit}
\ee
The zero mode $|0)$
has thus a vanishing norm in the limit $s\to -{1/2}\,$,
thereby recovering the results of the discrete series 
obtained in Section 2.2.

\section{Scalar field from complementary series}

Let ${\mathscr H}_\Omega$ be the one-dimensional Hilbert space
associated with the trivial representation $|\Omega)$
and $\mathscr H$ the Hilbert space carrying
the UIR of the preceding section. 
The Fock space ${\mathscr F}$ is constructed from 
these two spaces in the following manner
\be
	\label{Fock}
	{\mathscr F}=   {\mathscr H}_\Omega \oplus 
	\bigoplus_{n=1}^{\infty}{\mathscr H}^{\otimes_{\sst\rm sym} n}\,,
\ee
where the $n^{th}$ term ${\mathscr H}^{\otimes_{\sst\rm sym} n}$
is the representation obtained by taking the symmetrized
tensor product of $n$ copies of the UIR.
Then, the creation and annihilation operators are defined as
\ba
    & a^\dagger_m \vert \Omega)=\vert m)\,,
	\qquad a_m\vert \Omega)=0\,,\nn
	&[\,a_m\,,\,a^\dagger_n\,]=Q_m\,\delta_{m,n}\,,
\ea
where $Q_m$ is defined in eq.(\ref{Q_comp}).
The above relations can be used to define the 
 creation and annihilation operators of the position basis:
\ba
    & a^\dagger(\phi)\vert \Omega)=\vert \phi)\,,
	\qquad a(\phi)\vert \Omega)=0\,,\nn
    & [\,a(\phi)\,,\,a^\dagger(\phi')\,]=Q(\phi-\phi')\,.
\ea
The (reducible) representation of the dS group on the Fock space is obtained
from the irreducible representation by
\be
 	U=\sum_{m,n}\, {\cal U}_{mn} \, a^\dagger_m\,a_{n}\,,
\ee
with ${\cal U}_{mn}=\la m|\ {\cal U}\,|n\ra/Q_{m}Q_{n}=(m|\ {\cal U}\,n)/Q_{n}$. Each
$n$-particle sector of the Fock space is thus kept invariant under the
action of the group transformations. 

Using these operators, we shall now construct a local field, $\Phi(x)$ on
dS space. We shall use  the \emph{global} coordinate system $(t,\theta)$, where
$t$ is the time coordinate and varies from $-\infty$ to $+\infty$
and $\theta$ is an angle coordinate. In this coordinate system, the metric is
\be
   ds^2=-dt^2+\cosh ^2t\,d\theta^2.
\ee
Later we shall also use the conformal time for which
\be
	ds^2=\sin^{-2}\!\eta\,(-d\eta^2+d\theta^2)\,.
\ee 
The point with global coordinates $(0,0)$ is transported to
$(t,\theta)$  by the following {\it ordered} sequence: first one acts with
a boost generated by $K_1$ with parameter $t$ to reach $(t,0)$, and
then one reaches $(t,\theta)$ by a rotation with angle $\theta$. Notice also
that the point $(0,0)$ is left invariant by the boost generated by $K_2$.


 Taking into account the non-commuting character of $J$ and $K_1$ we have
 the central equation
\be
   \Phi(t,\theta)=e^{-i\theta J}e^{it K_1} \, \Phi(0,0) \, e^{-it K_1}e^{i\theta J}\,.
   \label{e1dS}
\ee
It determines the field operator $\Phi(t,\theta)$ from the operator $\Phi(0,0)$.
The other important equation follows from the fact that
$\Phi(0,0)$ must be invariant under transformations that leave the point $(0,0)$
invariant. In two dimensions only $K_2$ leaves $(0,0)$ invariant. Hence we 
impose
\be
   \label{e1dS2}
   [\,K_2,\,\Phi(0,0)\,]=0\,.
\ee 
Eq.(\ref{e1dS}) and eq.(\ref{e1dS2})
imply that $\Phi(t,\theta)$ obeys the Klein-Gordon equation with a mass squared term given by
$M^2=-s^2+1/4$, see I for the proof.
Besides covariance, we also impose that
 our field is free of interactions. Hence we write it as
 a linear combination of creation and annihilation operators.
 In the Fourier basis we have
\be
   	\Phi(0,0)=\frac{1}{\sqrt{2\pi}}
	\sum_{m=-\infty}^{\infty}\ c_m\,
   	a^\dagger_m + c^*_m\,a_m\,.
\ee

The field operator is thus fully determined by the $c$-number constants $c_m\,$.
The crucial point is that eq.(\ref{e1dS2}) is 
not satisfied by arbitrary superpositions. Indeed imposing eq.(\ref{e1dS2})
gives the following conditions:
\be
     {c_{m+2} \over c_{m}}=-\frac{m+s+\frac12}{m-s+\frac32}
     ={\gamma_{m+2}\over \gamma_{m}}\,,
\ee
with
\be
     \gamma_m=e^{im\frac\pi2}
     \frac{\Gamma\left(\frac{m}2+\frac s2+\frac14\right)}
     {\Gamma\left(\frac{m}2-\frac s2+\frac34\right)}
	\label{gamma}
\ee
The solution here depends on {\it two} complex 
 constants $c_0$ and $c_1$.
These determine all other coefficients by
\be
     c_{2m}=c_0{\gamma_{2m}\over \gamma_0}\,, \qquad
     c_{2m+1}=c_1{\gamma_{2m+1}\over \gamma_1}\,.
\ee

In order to find how to interpret 
the set of covariant field
operators characterized by 
$c_0$ and $c_1$, 
it is necessary to determine how these fields evolve in space-time.

\subsection{Field operator in position basis}

To determine the time evolution of the field,
we work in the position basis
\be
       \Phi(0,0)=\int_0^{2\pi}d\phi\, \Psi_0(\phi)\,a^\dagger(\phi)+\Psi^*_0(\phi)\,a(\phi)\,.
\ee
with
\be
	\Psi_0(\phi)= (\phi\vert\,\Phi(0,0)\,\Omega)\,.
\ee
From eq.(\ref{e1dS2}),
we get the following differential equation:
\be
      (\phi\vert\,K_2\,\Phi(0,0)\,\Omega)\,=\,
      -i\,\left\{\cos\phi\frac d{d\phi}-\left(s+\frac12\right)\sin\phi\right\}\Psi_0(\phi)\,=\,0\,.
\ee
The general solution is given by
\be
	\Psi_0(\phi)= A \, \Theta(\cos\phi) \,   (\cos\phi)^{-s-\frac12}+
	B \, \Theta(-\cos\phi) \,  (-\cos\phi)^{-s-\frac12}\,,
	\label{coef AB}
\ee
where $\Theta$ is the Heaviside step function and 
from the action of the dS transformation given in eq.(\ref{te}) and the covariance
of the field operator, we deduce the field operator at an arbitrary point in $dS_2$ as
\be
      \Phi(t,\theta)=\int_0^{2\pi}d\phi\, \Psi_{t,\theta}(\phi)\,a^\dagger(\phi)+\Psi^*_{t,\theta}(\phi)\,a(\phi)\,.
\ee
with
\ba
	\Psi_{t,\theta}(\phi)&=&
     	A\,\Theta(\cosh t\cos(\phi-\theta)+\sinh t)\,(\cosh t\cos(\phi-\theta)+\sinh t)^{-s-\frac12}+
	\nn&&+\,
	B\,\Theta(-\cosh t\cos(\phi-\theta)-\sinh t)\,(-\cosh t\cos(\phi-\theta)-\sinh t)^{-s-\frac12}\,.
\ea
It is important to notice that although 
we got a first order differential equation,
being singular at $\phi_{\pm}=\theta \pm \arccos(-\tanh t)$, its solution
depends on two complex numbers (which can then be related to $c_0$ and $c_1$).
Notice also that the interval 
$\mathop{[\phi_-,\phi_+]}$ corresponds to the region of space which
is in the causal infinite past of $(t,\theta)$ (see, Fig.\ref{fig}).
 \EPSFIGURE{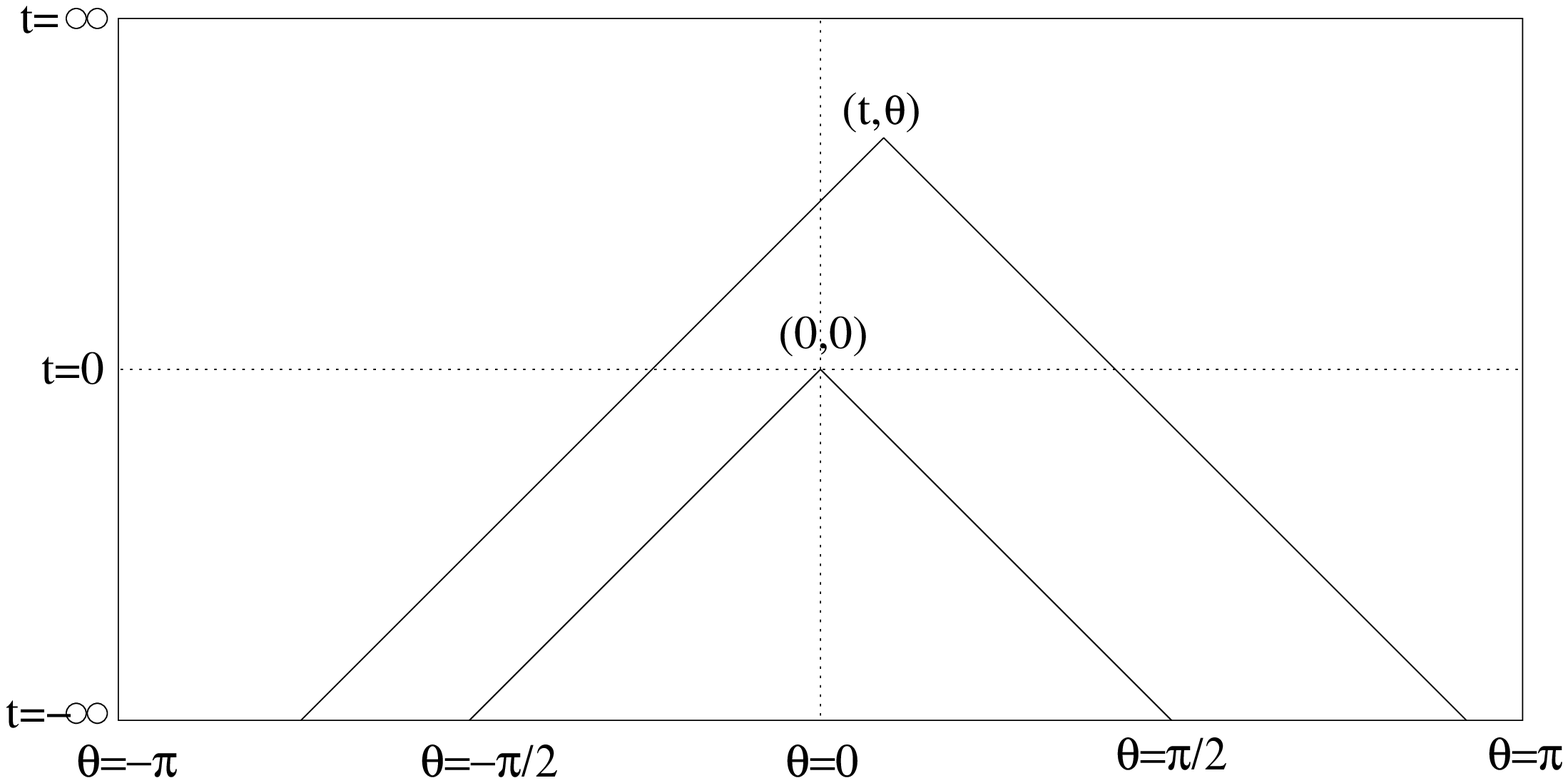,width=9cm}{\label{fig}
	The Carter-Penrose diagram 
	of the two dimensional dS space.
	We have represented the causal past of the point $(t,\theta)$.}
 From this we learn that the two-folded degeneracy of covariant
 field operators is deeply related to the causal structure of 
 de Sitter space.

In the sequel it will be also useful to use 
the {\it regular} holomorphic and anti-holomorphic combinations:
\be
	\Psi_{t,\theta}(\phi)=
     	C\,(\cosh t\cos(\phi-\theta)+\sinh t-i\epsilon)^{-s-\frac12}+
	\,D\,(\cosh t\cos(\phi-\theta)+\sinh t+i\epsilon)^{-s-\frac12}\,.
	\label{state CD}
\ee
Writing, as usual, the field operator as
\be
	\Phi(t,\theta)=\frac{1}{\sqrt{2\pi}}\sum_{m=-\infty}^{\infty}
	c_m(t)\,e^{-im\theta}\,a^\dagger_m+c_m^*(t)\,e^{im\theta}\,a_m\,.
\ee
The coefficient $c_m(t)$ are thus given by
\ba
     c_m(t)&=&\sqrt{2\pi}\,(m\vert\,\Phi(t,0)\,\Omega)
     =\int_0^{2\pi} d\phi\,e^{-im\phi}\,\Psi_{t,0}(\phi)
     \nonumber\\
	&=&\Gamma(-s+{\st\frac12})\left(\frac{2\pi}{\cosh t}\right)^{\frac12}
       \Big[
       (C+D)\,{\rm P}^{s}_{\sst m-{1\over 2}}(-\tanh t)+
	\nn
	&&
       +\left(C\,e^{-i\pi(s+\frac{1}{2})}+D\,e^{i\pi(s+\frac{1}{2})}\right)(-1)^m\,
       {\rm P}^{s}_{\sst m-{1\over 2}}(\tanh t)\Big]\,,
	\label{coeff}
\ea
and from this, we get the relation between $C, D$ and $c_0, c_1\,$:
\ba
      \frac{\Gamma\left(-\frac s2+\frac34\right)}{\sqrt{\pi}\,\Gamma\left(-\frac s2+\frac14\right)}
      \, c_0&=&
       \left(1+(e^{-i\pi})^{-s-\frac12}\right)C+\left(1+(e^{i\pi})^{-s-\frac12}\right)D\,,
 	\nonumber\\
    \frac{\Gamma\left(-\frac s2+\frac54\right)}{\sqrt{\pi}\,\Gamma\left(-\frac s2+\frac34\right)}
      \, c_1&=&
       \left(1-(e^{-i\pi})^{-s-\frac12}\right)C+\left(1-(e^{i\pi})^{-s-\frac12}\right)D\,.
	\label{c01 CD}
\ea

\subsection{Commutation relations}

In this subsection we show that the covariant fields, 
solutions of eq.(\ref{e1dS2}),
automatically obey canonical commutation relations
 up to an overall constant we
shall determine.

Since $\gamma_m=\gamma_{-m}$, we have $c_{m}=c_{-m}$ for all $m$
which implies that any covariant field is 
invariant under parity: $\theta \to - \theta$. 
In turn this implies the following commutation relation
\be
     [\,\Phi(0,\theta),\,\Phi(0,0)\,]=0\,.
\ee
See I for the details of the proof. 
Next, we consider the commutator of the field with its conjugate momentum
$\Pi=\partial_t\Phi\,$. Using
\be
     -i\,\partial_t \Phi(0,0)=[\,K_1,\,\Phi(0,0)\,]=
	\sum_{m=-\infty}^{\infty}\frac{i}{\sqrt{2\pi}}\left(m+s+\frac12\right)c_m\,a_{m+1}^\dagger+{\rm h.c.}
\ee
we obtain
\be
     h(\theta)=i\,[\,\Pi(0,0)\,,\Phi(0,\theta)\,]=
	\sum_{m=-\infty}^{\infty}\frac{i}{2\pi}\left(m+s-\frac12\right)
     \,c_m^*c_{m-1}\,Q_m\,e^{im\theta}+{\rm c.c.}
\ee
Now we use
\be
     \left(m+s+\frac12\right)\gamma^*_{m}\gamma_{m-1}\,Q_m=-i\,2^{-2s+1}\,
     \frac{\Gamma\left(m_0+s+\frac12\right)}{\Gamma\left(m_0-s+\frac12\right)}
     \la m_0|m_0\ra\,,
\ee
to get
\be
     h(\theta)=
     {\rm Re} \left(\frac{ c_0\,c_1^*} {\gamma_0 \,\gamma_1^*}\right)
     2^{-2s+1}\,\frac{\Gamma\left(m_0+s+\frac12\right)}{\Gamma\left(m_0-s+\frac12\right)}
     \la m_0|m_0\ra\,\delta(\theta)\,.
\ee 
The canonical commutation relations are thus satisfied
if the coefficient of the Dirac function $\delta(\theta)$ is one. 
From this and the fact that $\gamma_0 \gamma_1$ is 
 purely imaginary, as can be seen from eq.(\ref{gamma}),  we get
 \be
 	 c_1^*c_0-c_1c_0^*=
	i\frac{\Gamma\left(s-\frac12\right)}
     	{\Gamma\left(-s+\frac12\right)}
	\frac{\Gamma\left(m_0-s+\frac12\right)}
     	{\Gamma\left(m_0+s+\frac12\right)}{1\over \la m_0|m_0\ra}\,.
	\label{canon c01}
\ee
In terms of $C$ and $D$, this condition, using eq.(\ref{c01 CD}),
reads
\be
	|D|^2-|C|^2=
	\frac{\Gamma\left(s+\frac12\right)}
     	{\Gamma\left(-s+\frac12\right)}
	\frac{\Gamma\left(m_0-s+\frac12\right)}
     	{\Gamma\left(m_0+s+\frac12\right)}{1\over \la m_0|m_0\ra}\,
     	\frac1{8\pi\,\cos\pi s} \equiv N^2 \,.
	\label{canon CD}
\ee
This quadratic form is invariant under
SU$(1,1)$ transformations.

\subsection{Hadamard requirement and Bunch-Davies vacuum}

To understand the meaning of the set of covariant canonical 
fields, $\Phi_{c_0,c_1}(t,\theta)$, 
we consider the two-point function evaluated in the vacuum 
\be
	G_{c_0,c_1}(t,\theta;t',\theta') \equiv 
	(\Omega|\,\Phi_{c_0,c_1}(t,\theta)\,\Phi_{c_0,c_1}(t',\theta')\ \Omega)\,.
	\label{2d 2pt fn}
\ee
This observable, which parametrically depends on $c_0$ and $c_1$,
can then be compared with the corresponding quantity 
obtained in the usual quantization of scalar fields
on de Sitter space.
This correspondence is most easily achieved 
when requiring that $G(0,\theta;0,0)$ possesses 
the Hadamard behavior  in the coincidence point limit.
This limit is governed by the high $m$ behavior of the coefficients $c_m$. 
It coincides with that of flat space if
\be
     \frac{1}{2\pi}|c_m|^2\,Q_m\underset{|m|\to\infty}{\approx}{1\over 4\pi|m|}\,.
\ee
The asymptotic behaviors of $|\gamma_m|^2$ and $Q_m$ are given by
\be
     |\gamma_{m}|^2\underset{|m|\to\infty}{\approx}
     \left(\frac m2\right)^{2s-1}\,,
     \qquad
     Q_m\underset{|m|\to\infty}{\approx}
     m^{-2s}\,
	\frac{\Gamma\left(m_0+s+\frac12\right)}{\Gamma\left(m_0-s+\frac12\right)}\,\la m_0|m_0\ra\,.
\ee
The Hadamard condition thus gives
\be
     \frac{|c_0|^2}{|\gamma_0|^2}=\frac{|c_1|^2}{|\gamma_1|^2}
     =	\frac{\Gamma\left(m_0-s+\frac12\right)}{\Gamma\left(m_0+s+\frac12\right)}\,
	\frac{2^{2(s-1)}}{\la m_0|m_0\ra}\,.
\ee
If we  combine it with the canonical normalization,  eq.(\ref{canon c01}), we get a unique
solution we shall call Bunch-Davies (BD) \cite{BD}

\be
     	\frac{c_0^{\sst\rm BD}}{\gamma_0}
	=\frac{c_1^{\sst\rm BD}}{\gamma_1}
	=\sqrt{
	\frac{\Gamma\left(m_0-s+\frac12\right)}{\Gamma\left(m_0+s+\frac12\right)}\,
	\frac{2^{2(s-1)}}{\la m_0|m_0\ra}}\,.
	\label{BD coeff}
\ee
Using eq.(\ref{canon CD}), the BD condition simply reads
\be
     	C^{\scriptscriptstyle\rm BD}=0\,,
	\qquad 
	D^{\scriptscriptstyle\rm BD}= N\,.
\ee
These equations show that the BD solution can be univoquely 
characterized by the canonical normalization and the
anti-holomorphic behavior of the function $\Psi_0$ of eq.(\ref{state CD}).


Using the above as the reference solution, the moduli space 
of canonical fields can be
parametrized by
\be
     c_0=c_0^{\scriptscriptstyle\rm BD}(\cosh\alpha+e^{i\beta}\sinh\alpha)\,,
     \qquad
     c_1=c_1^{\scriptscriptstyle\rm BD}(\cosh\alpha-e^{i\beta}\sinh\alpha)\,.
     \label{para_BD}
\ee
The moduli space is SU$(1,1)/$U$(1)$. As explained for the principal series in
I, it corresponds to the the space of Bogoliubov transformations 
which preserve de Sitter invariance and which relate two alpha-vacua \cite{Allen:1985ux}.
Let us denote by $\Phi_{\sst\mathrm{BD}}^+(x)$ the part of the BD field operator 
 which contains only creation operators.
Then, repeating the steps 
 of Section 5 of \cite{Joung:2006gj},
 the general field can be expressed as
\be
	\Phi_{\alpha,\beta}(x)=\cosh\alpha\,\Phi_{\sst\mathrm{BD}}^+(x)+e^{i\beta}\sinh\alpha
	\,\Phi_{\sst\mathrm{BD}}^+(\bar x)+{\rm h.c.}
	\label{gener}
\ee
 where $\bar x=(\pi-\eta,\theta+\pi)$ is the antipodal point to $x=(\eta,\theta)$.

We have reached the interpretation of our SU$(1,1)/$U$(1)$ set of covariant
and canonical fields $\Phi_{\alpha,\beta}$.
They provide an alternative description of the fact the canonical
quantization of fields on de Sitter space supplemented 
by the requirement that the vacuum state
be de Sitter invariant leads to a SU$(1,1)/$U$(1)$ class of 
de Sitter invariant two-point functions. In the standard approach
these two-point functions are viewed as expectation values built with
{\it the} field operator in a class of states, the alpha-vacua.
Here instead, the two-point functions are given by eq.(\ref{2d 2pt fn}),
which is the expectation values of different field operators 
evaluated the unique vacuum
state $|\Omega)$ which carries the trivial representation. 

\subsection{{\emph{in}} and {\emph{out}} vacuum}

One of the important physical difference between the fields
 in the principal and complementary series is 
their behavior at large $t$. In this limit the Klein-Gordon equation reads
\be
	(\partial_t^2+\partial_t+M^2)\Phi=0\,.
\ee
Its solutions are a combination of $e^{-t/2}\,e^{\pm i\mu t}$ with
\be
\mu=\sqrt{M^2-{1\over 4}}\,.
\ee
The principal series is characterized by $M^2>{1\over 4}$
leading to solutions with oscillatory behavior and allowing the
determination of $in$ and $out$ vacua as positive (proper) frequency modes
in the remote past and future respectively \cite{Mottola:1984ar}.
For the complementary series we have instead $\mu=i\,s$ purely imaginary.
Therefore the asymptotic solutions are exponentially decreasing in proper time,
and the definition of $in$ and $out$ vacua no longer applies.
The same problem arises for $M=0$ where $\mu=-i/2\,$.
In this case however, the conformal invariance of the Klein-Gordon equation
leads to  modes varying as
$e^{\pm i m\eta}$. The positive conformal
frequency modes define the conformal vacuum
which coincides with
 $in$ and $out$ asymptotic vacua since there is no frequency mixing.
Here, we show that we can define  $in$ and $out$ vacua
for the complementary series as positive frequency solutions with respect
to a time coordinate which interpolates
between the proper time and the conformal time.

In the remote past, the  coefficient $c_m(t)$ has the asymptotic behavior:
\be
	c_m(t)\underset{t\to-\infty}{\propto}
	e^{\left(s+\frac12\right)t}
	\left\{1+i\,\omega^{\sst\rm in}_m\,
	\left(2\,e^{-2st}\right)\right\},
\ee
with
\be
	\omega^{\sst\rm in}_m=
	{\Gamma(1+s)\,\Gamma(m-s+{1\over 2})\over 2\,\Gamma(1-s)
	\,\Gamma	(m+s+{1\over 2})}\,
	\frac{-\sin\pi s\,+i(\cos\pi s \cosh 2\alpha-\sin \beta\sinh 2\alpha)}
	{\cosh2\alpha-\sin(\pi s+\beta)\,\sinh2\alpha}.
	\label{in_fq}
\ee
Let us factorize the overall decreasing term(in $|t|$) 
and define a new  time coordinate which in the infinite past is related to $t$ as by
\be
	\eta^{\sst\rm }_s=2\,e^{-2st}.
	\label{redef time}
\ee
It goes to zero in the remote past and reduces to the conformal time
in this limit for $s={-1/2}\,$.
 The last factor of $c_m(t)$ can be written as a plane wave 
to the first order in $\eta_s$:
\be
	e^{i\,\omega^{\sst\rm in}_m\,
	\eta^{\sst\rm }_s} \,
	+ {\cal O}({\eta^{\sst\rm }_s}^{2})\,.
	\label{in plane}
\ee
The $in$ vacuum with respect to the $\eta_s$ 
time coordinate is now defined by 
$ \omega^{\sst\rm in}_m$ real and positive.
The condition for  $\omega^{\sst\rm in}_m$ to be real is
\be
	\tanh2\alpha^{\sst\rm in}\,
	\sin\beta^{\sst\rm in}=\cos\pi s\,,\label{in}
\ee
and the resulting value of  $\omega^{\sst\rm in}_m$
 is given by
\be
	\omega^{\sst\rm in}_m
	=
	{\Gamma(1+s)\,\Gamma(m-s+{1\over 2})
	\over 2\,\Gamma(1-s)\,\Gamma(m+s+{1\over 2})}
	{\sqrt{\sin^2\beta-\cos^2\pi s}\over \cos(\pi s+\beta)},
\ee
which is positive for $\beta$ verifying eq.(\ref{in}) and so 
$\pi(1/2+s)<\beta<\pi(1/2-s)\,$.
Notice that we get a family of $in$ vacua since $\beta$ is an arbitrary angle
in the interval $\,\mathop{]\pi(1/2+s),\pi(1/2-s)[}\,$.
In the limit $s\to -1/2$, the frequency $\omega^{\sst\rm in}_m$
reduces as it should to $|m|$ which is the conformal frequency.
The large $m$ limit of  $\omega^{\sst\rm in}_m$
varies as $m^{-2s}$ which has the expected behavior when the complementary
series joins the principal one that is for $s\to 0$.

The large future limit is given by
\be
	c_m(t)\underset{t\to\infty}{\propto}
	e^{-\left(s+\frac12\right)t}
	\left\{1+i\,\omega^{\sst\rm out}_m\,
	\left(-2\,e^{2st}\right)\right\},
\ee
with
\be
	\omega^{\sst\rm out}_m=
	{\Gamma(1+s)\,\Gamma(m-s+{1\over 2})\over 
	2\,\Gamma(1-s)\,\Gamma(m+s+{1\over 2})}\,
	\frac{\sin\pi s\,+i(\cos\pi s \cosh 2\alpha+\sin \beta\sinh 2\alpha)}
	{\cosh2\alpha+\sin(\pi s-\beta)\,\sinh2\alpha}.
	\label{out_fq}
\ee
We demand that the time coordinate $\eta_s$ is the remote future
has an asymptotic expression in terms of $t$ given by
\be
	\eta_s=\pi-2e^{2st},
	\label{redef time out}
\ee
so that $c_m(t)$ has an $\eta_s$ dependence given by $e^{i\,\omega^{\sst\rm out}_m\, \eta_s}\,$.
The \emph{out} vacuum is such that $\omega^{\sst\rm out}_m$ is real and positive.
The condition for $\omega^{\sst\rm out}_m$ to be real is 
\be
	\tanh2\alpha^{\sst\rm out}\,\sin\beta^{\sst\rm out}=-\cos\pi s\,.
	\label{out}
\ee
and the resulting value of  $\omega^{\sst\rm out}_m$
 is given by
\be
	\omega^{\sst\rm out}_m
	=
	{\Gamma(1+s)\,\Gamma(m-s+{1\over 2})
	\over 2\,\Gamma(1-s)\,\Gamma(m+s+{1\over 2})}
	{\sqrt{\sin^2\beta-\cos^2\pi s}\over \cos(\pi s-\beta)},
\ee
which is positive for $\beta$ verifying eq.(\ref{out}) and so 
$-\pi(1/2-s)<\beta<-\pi(1/2+s)$.
Notice that we get a family of $out$ vacua since $\beta$ is an arbitrary angle
in the interval $\,\mathop{]\!-\!\pi(1/2-s),-\pi(1/2+s)[}\,$.

An explicit expression for the time coordinate $\eta_s$ which gives
the required asymptotic behavior for large and small $t$ is 
\be
	\tan{\eta_s\over 2}=e^{-2st}\,.
\ee
It reduces to the conformal time in the limit $s\to -{1/2}\,$.

The mean number of \emph{out} quanta of momentum $m$ in \emph{in} vacuum is
\be
	\bar{n}_{\sst\rm out/in}=
	\left|
	\cosh\alpha^{\sst\rm out}\,\sinh\alpha^{\sst\rm in}\,
	e^{i\beta^{\sst\rm in}} -
	\cosh\alpha^{\sst\rm in}\,\sinh\alpha^{\sst\rm out}\,
	e^{i\beta^{\sst\rm out}}
	\right|^2
\ee
If we compute this between time-reversal \emph{in} and \emph{out} vacua with
$\alpha^{\sst\rm out}=\alpha^{\sst\rm in}$,
$\beta^{\sst\rm out}=-\beta^{\sst\rm in}$, then we get
\be
	\bar{n}_{\sst\rm out/in}
	=\cosh^2 (2\alpha^{\sst\rm in})\,\cos^2\pi s
	=\frac{\sin^2\beta^{\sst\rm in}\,\cos^2\pi s}{\sin^2\beta^{\sst\rm in}-\cos^2\pi s}
\ee
Notice that when $\beta^{\rm in}=\pi/2$ the number of created quanta is minimal
and is given by $\cot^ 2 \pi s$. It is easy to show that this is
also the minimal number of created quanta when considering
all \emph{in} and  \emph{out} vacua.

In conclusion we have defined for the complementary series
 $in$ and $out$ vacua 
which belong to the SU$(1,1)/$U$(1)$ moduli space of dS invariant vacua.
Even though they are not unique for a given value of $s$
they fulfill the criterion of being associated with positive frequency modes
with respect to some asymptotic time parameter. The other unusual property is that this
parameter now depends
on $s$. These states could be relevant in certain inflationary
models when considering nearly massless fluctuation modes with a value of $s$
belonging to the complementary series but close to the discrete series
which represents in four dimensions the massless minimally coupled field. 

\section{Massless scalar field from discrete series}

Consider the UIR $D^+_0$ of Section 2.2 and
expand $\Phi_+(0,0)$ as
\be
      \Phi_+(0,0)=\frac{1}{\sqrt{2\pi}}\sum_{m=1}^{\infty}c_m\,a^\dagger_m+c_m^*\,a_m\,,
\ee
where
\be
       [\,a_m\,,\,a^\dagger_n\,]=\frac{\la m_+|m_+\ra}{m_+}\,m\,\delta_{m,n}\,,
       \qquad m\,,n\,,m_+\in{\mathbb Z}_{>0}\,.
\ee
The locality condition $[\,K_2\,,\,\Phi_+(0,0)\,]=0$ gives
\be
      c_{2m}=0\,,\qquad (2m+1)c_{2m+1}=(-1)^m\,c_1\,.
\ee

The commutator at equal time can now be readily calculated
\ba
      [\,\Phi_+(0,\theta)\,,\,\Phi_+(0,0)\,]
      &=&
	\frac{\la m_+|m_+\ra}{m_+}\,\frac{|c_1|^2}{2\pi}\,\sum_{m=0}^{\infty}
      \frac{e^{i(2m+1)\theta}-e^{-i(2m+1)\theta}}{2m+1}
	\nonumber\\
	&=&
	i\,\frac{\la m_+|m_+\ra}{m_+}\,|c_1|^2\,{\rm sgn}(\theta)\,.
\ea
where
${\rm sgn}(\theta)$ is $2\pi$-periodic, odd and equals $+1$ for
$0<\theta<\pi$ and $-1$ for $\pi<\theta<2\pi$. The field is not a canonical one
since the commutator at equal times does not vanish. The above commutation
relation implies that
\be
	[\,\partial_\theta\Phi_+(0,\theta)\,,\,\Phi_+(0,0)\,]
     	=2 i\,\frac{\la m_+|m_+\ra}{m_+}\,|c_1|^2\,{\delta}(\theta)\,.
	\label{ccr ads}
\ee
Using the fact that the time derivative at $t=0$ is 
\be
      \partial_t\Phi_+(0,0)=\frac{1}{\sqrt{2\pi}}\sum_{m=0}^{\infty}
     \,(-1)^{m+1}\left(c_1\,a^\dagger_{2m+2}-c_1^*\,a_{2m+2}\right)\,.
\ee
we deduce that $\partial_t\Phi(0,0)$ and $\Phi(0,\theta)$ commute.
Hence, $\Phi$ is not a canonical field. 

In this we recover the fact that 
it is impossible to construct a canonical and covariant massless 
field on dS space.
This is indeed in agreement with the result of Allen \cite{Allen:1985ux},
who started with a massless canonical field and showed that
it has no dS invariant two-point function.
Here instead we have a well defined two-point function
but we have lost the canonical commutation relations.

\subsection{Non canonical massless field}

In spite of the fact our field is non-canonical it is worth
to further analyze its properties.

First, it obeys the conformally invariant
equation $(\partial_\eta^2-\partial^2_\theta)\Phi=0$.
This allows a simple determination of the
 field operator at an arbitrary point
 from the operator and its
time derivative at $\eta=\pi/2$ and $\theta =0$. A simple calculation yields
\be
      \Phi_+(\eta,\theta)=\frac{c_1}{\sqrt{2\pi}}\sum_{m=1}^{\infty}
     \frac{\sin(m\eta)}m \left(e^{-im\theta}\,{a^\dagger_m}
      +e^{im\theta}\,{a_m}\right)\,.
	\label{nc field}
\ee

Second, its Wightman  function
$G(\eta,\theta\,;\,\eta',\theta')=(\Omega|\,\Phi_+(\eta,\theta)\,\Phi_+(\eta',\theta')\,\Omega)$
is well defined, thanks to the absence of the zero mode. Explicitly, one finds:
\be
      G(x;x')
      =\frac{\la m_+|m_+\ra}{m_+}\,\frac{|c_1|^2}{8\pi}\Bigg[\,
      \ln\left|\frac{1+Z(x;x')}{1-Z(x;x')}\right|
      +i\pi\,{\rm sgn}(\theta-\theta')\,\Theta\left(1-Z^2(x;x')\right)
      \,\Bigg]\,.
	\label{nc 2pt fn}
\ee
where
\be
      Z(\eta,\theta\,;\,\eta',\theta')
      =\frac{\cos(\theta-\theta')-\cos\eta\,\cos\eta'}{\sin\eta\,\sin\eta'}
      =X^\mu X'_\mu\,.
\ee
This two-point function is dS invariant as can be easily seen from 
the embedding of $dS_2$ in a flat three dimensional space.
The product $Z^{(1,2)}(x;x')=X^\mu X'_\mu$ is an invariant which is symmetric with respect to
the exchange of the two spacetime points.
It is related to the invariant distance $\sigma(x;x')$ between the two spacetime
points by $Z(x;x')=\cos\left\{\sigma(x;x')\right\}$.
When the vector $\epsilon_{\mu\nu\rho}X^\nu X'^\rho$ 
is timelike, an antisymmetric invariant is given by the sign of the time
component. In this case we recover 
the function ${\rm sgn}(\theta-\theta')\,\Theta(1-Z(x;x')^2)$ 
introduced above.\footnote{
When $(X^\mu-X'^\mu)(X_\mu-X'_\mu)$
is negative, the antisymmetric invariants
are given by the sign of the time component $X^0-X'^0$ 
which is also the sign of the difference $t-t'$ or $\eta-\eta'$ when $Z(x;x')>1$.
Explicitly one has
\be
      s^{(1,2)}(x;x')={\rm sgn}(\eta-\eta')\,\Theta(Z(x;x')-1)\,.
        \label{s12}
\ee
When changing $X'^\mu$ to $-X'^\mu$, that is, when replacing $x'$ 
by its antipodal point $x'_A$, gives the other invariant
\be
        s^{(1,2)}_A(x;x')=s^{(1,2)}(x;x'_A)={\rm sgn}(\eta+\eta'-\pi)\,\Theta(-Z(x;x')-1)\,.
\ee
}

Third, our field operator is closely related to the 
two-dimensional fields used in string theory \cite{Green:1987sp}.
Indeed, up to a normalization convention,
our field coincides with that describing 
the coordinate of an open string with
Dirichlet boundary conditions at both ends,
and with the spatial 
and temporal coordinates interchanged:
$\Phi_+(\eta,\theta)=X(\tau=-\theta,\sigma=\eta)\,$.
At this point it 
is worth to notice that the interchange of the spatial and temporal 
coordinates gives a canonical massless field defined on $AdS$.
In fact under this exchange, one recovers that the creation operators
in eq.(\ref{nc field}) have only positive frequencies.
Moreover in that case, the canonical commutation relations fixes the
normalization of $c_1$ in eq.(\ref{ccr ads}).

Fourth, as in string theory, because
of the conformal invariance, our field is covariant under a larger group of transformations. 
The generators of this group are
\be
	L_m=
	\sum_{n=1}^{\infty}a_n^{\dagger}\,a_{m+n}+
	\frac12\sum_{n=1}^{m-1}a_n\,a_{m-n}\,,
	\qquad 
	L_{-m}=L_m^\dagger\,,
	\qquad
	m\ge0\,,
	\label{virasoro}
\ee
and they act on the field as
\ba
	[\,L_m\,,\, \Phi_+(\eta,\theta)\,]&=&
	i\left(e^{-im (\theta+\eta)}\partial_{\theta+\eta}+
	e^{-im (\theta-\eta)}\partial_{\theta-\eta}\right)\Phi_+(\eta,\theta)
	\nn&=&e^{-im \theta}
	\left(\sin m\eta\,\partial_\eta+i\cos m\eta\,\partial_\theta\right)
	\Phi_+(\eta,\theta)\,,
	\label{vira act}
\ea
and satisfy the Virasoro algebra with central charge $c=1\,$:
\be
	[\,L_m\,,\,L_n\,]=(m-n)\,L_{m+n}+\frac{1}{12}(m^3-m)\delta_{m,-n}\,.
\ee
Notice that the generators of the dS group correspond to 
$L_{\pm1}=\mp i\,K_\pm$ and $L_0=J$.
These three generators do not 
contain pair of creation or annihilation operators, 
as can be seen from eq.(\ref{virasoro}).

In spite of these well defined properties, 
the non-canonical character of our field shows up when considering
the normal ordered operator of the energy-momentum tensor 
$T_{\mu\nu}=\;:\partial_\mu\Phi\partial_\nu\Phi-\frac{1}{2}g_{\mu\nu}\partial^\rho\Phi\partial_\rho\Phi:$
and the Hamiltonian operator: 
\be
	H= \int_0^{2 \pi}{d\theta}\;T_{\eta\eta}=\int_0^{2 \pi}{d\theta}\,\left(T_{++}+T_{--}\right)\,,
\ee
where the subscripts $\pm$ are with respect to the light cone variables $x^\pm = \eta  \pm \theta$.
$T_{++}$ and $T_{--}$ give also the generators of Virasoro algebra, $L_m\,$, as
\be
	\frac{c^2}{2}\,L_m=
	\int_{0}^{2\pi}dx^+\,e^{-imx^+}T_{++}(x^+)
	=\int_{0}^{2\pi}dx^-\,e^{imx^-}T_{--}(x^-)\,,
\ee
so one finds immediately $H=c^2\,J$ and 
the Hamiltonian $H$ generate rotations rather than time translations.
In fact calculating the $\theta$-translation generator in
the Hamiltonian formalism as
\be
	P_\theta=\int_0^{2 \pi}{d\theta}\;T_{\eta\theta}=\int_0^{2 \pi}{d\theta}\,\left(T_{++}-T_{--}\right)\,,
\ee
we  get  identically $P_\theta=0\neq J$.
This is due to the interchange of the role of time and space
in the canonical commutation relation. 
We consider also
the transition amplitude of a detector \cite{Unruh:1975gz}
with resonance frequency $E$, coupled linearly to $\Phi$,
and located at constant $\theta$.
This amplitude is proportional to
\be
      A(E)=\int_{-\infty}^\infty dt
      \, e^{-iEt}\,G(t,\theta;0,\theta)
      =\frac{|c|^2}{4E}\,\tanh\left(\frac\pi2E\right).
\ee
It is even in $E$ so the
detector has the same probability of loosing energy or gaining the
same amount of energy. It cannot reach thermal equilibrium
since the final steady state will be characterized by equally 
populated states (infinite temperature).

 \subsection{Parity invariant massless field}

It should be noticed that our field is not invariant under parity.
In what follows we shall construct a parity invariant field
and show that the above sicknesses  are not cured by this new field. 

Consider first the scalar field constructed from the other discrete
series $D^-_0$. Its expansion reads
\be
      \Phi_-(\eta,\theta)=\frac{c_{-1}}{\sqrt{2\pi}}\sum_{m=-\infty}^{-1}
      \frac{\sin(m\eta)}m\left(e^{-im\theta}\,a^\dagger_{m}+ e^{im\theta}\,a_{m}\right).
\ee
where
\be
	[\,a_{m}\,,\,a^\dagger_{n}\,]=
	\frac{\la m_-|m_-\ra}{m_-}\,
	m\,\delta_{m,n}\,,\qquad m,\,n,\,m_-\in{\mathbb Z}_{<0}\,.
\ee
The generators of a new Virasoro algebra $\tilde L_m$'s are given by
\be
	\tilde L_m=\sum_{n=1}^{\infty}a_{-n}^{\dagger}\,a_{-m-n}+
	\frac12\sum_{n=1}^{m-1}a_{-n}\,a_{-m+n}\,,
	\qquad 
	\tilde L_{-m}=\tilde L_m^\dagger\,,
	\qquad
	m\ge0\,.
\ee
The generators of the original dS group are 
$\tilde L_{\pm1}=\mp i\,K_\mp$ and $\tilde L_0=-J$.
The Virasoro algebra generated by $\tilde L_m$ 
have also a central charge $c=1\,$, and
act on the field as
\ba
	[\,\tilde L_m\,,\, \Phi_-(\eta,\theta)\,]&=&
	-i\left(e^{im (\theta+\eta)}\partial_{\theta+\eta}
	+e^{im (\theta-\eta)}\partial_{\theta-\eta}\right)
	\Phi_-(\eta,\theta)
	\nn&=&e^{im \theta}
	\left(\sin m\eta\,\partial_\eta-i\cos m\eta\,\partial_\theta\right)
	\Phi_-(\eta,\theta)\,.
\ea
Its two-point function is given by the complex conjugated of that
of positive series (\ref{nc 2pt fn}), up to a normalization constant.

We can now construct a parity invariant field
by taking the sum of the two previous fields with $c_{1}=c_{-1}=c$, and
with $\la m_+|m_+\ra/m_+=-\la m_-|m_-\ra/m_- $ which
we put to one for $m_\pm=\pm1\,$. Explicitly, one has
\ba
      \Phi(\eta,\theta)&=&\Phi_+(\eta,\theta)+\Phi_-(\eta,\theta)
	\nonumber\\
	&=&\frac{c}{\sqrt{2\pi}}\sum_{m=1}^{\infty}  {\sin(m\eta) \over m}
      \left\{e^{-im\theta}\left(a^\dagger_m+a_{-m}\right)
      +e^{im\theta}\left(a^\dagger_{-m}+a_{m}\right)\right\}\,.
	\label{pi nc field}
\ea
The resulting field
has a vanishing commutator for all spacetime points
\be
      [\,\Phi(\eta,\theta)\,,\,\Phi(\eta',\theta')\,]=0\,.
\ee
It therefore behaves as a classical field.

In addition, this is reflected in the 
Virasoro algebra generated by $\mathsf L_m$
induced from the $L_m$ of $\Phi_+$ and $\tilde L_m$ of $\Phi_-$:
\be
	\mathsf L_m=L_m-\tilde L_{-m}\,.
\ee
which has a vanishing central charge.
Even though the field has both positive and negative modes which correspond
to left-moving and right-moving string oscillation modes, it has only
one set of Virasoro algebra $\mathsf L_m$.
The other set of Virasoro algebra
generated by $\tilde{\mathsf L}_m=L_m+\tilde L_{-m}$ does not leave the 
field covariant. 
Notice finally that the Wightman function is twice the
symmetrical part of that of eq.(\ref{nc 2pt fn}). It is therefor real.

\section{Massless limit of massive field}

Given that we obtained a set of canonical fields when dealing with the
complementary series and no canonical field for the discrete series, 
it is interesting to consider the massless limit
from the complementary series: $\varepsilon=s+1/2\to 0$.

From eq.(\ref{coeff}) and eq.(\ref{BD coeff}),
the BD coefficients are given by
\be
      c_m^{\sst\rm BD}(\eta)=
     \frac{1}{\sqrt{2}}\left\{
      \begin{array}{ll}
      \varepsilon^{-1}+\ln(2\sin\eta)-\frac12+i\left(\eta-\frac\pi2\right)
	+{\cal O}(\varepsilon)\;&;\;m=0\\
      {|m|}^{-1}\,e^{i|m|\eta}+{\cal O}(\varepsilon)\;&;\;m\neq 0
      \end{array}
      \right.\,,
	\label{mdb}
\ee
where we chose the  normalization $m_0=1=\la 1|1\ra\,$. 
The commutation relations and the scalar product behave as
\be
	[\,a_m\,,\,a_m^\dagger\,]=\la m|m\ra
	=\left\{
      \begin{array}{ll}
     \varepsilon(1-\varepsilon)^{-1}\;&;\;m=0\\
	|m|+{\cal O}(\varepsilon)\;&;\;m\neq0
      \end{array}\right..
\ee
From eq.(\ref{mdb}), we deduce the behavior of the BD field 
\ba
      \Phi_{\sst\rm BD}(\eta,\theta)&=&
	{1\over 2\sqrt{\pi}} 
      \left[\left(\frac1\varepsilon+\ln(2\sin\eta)-\frac12\right)\left(a_0^{\dagger}+a_0\right)+
      i\left(\eta-\frac\pi2\right)\left(a_0^{\dagger}-a_0\right)\right]+
	\nonumber\\
	&&
      +\,{1\over 2\sqrt{\pi}}\sum_{m\neq0}\frac{1}{|m|}
      \left(e^{i|m|\eta-im\theta}\,a^\dagger_m
      +e^{-i|m|\eta+im\theta}\,a_m\right)\,+\,{\cal O}(\varepsilon)\,.
\ea
In the limit $\varepsilon\to 0$, the zero mode  
contains a $\varepsilon^{-1}$ divergence whereas  the commutation relation of 
$a_0$ and $a_0^\dagger$ vanishes as $\varepsilon$.
 The two combinations which appear in the zero mode are
\be
      q=\,\frac{1}{2\sqrt{\pi}\,\varepsilon}\left(a_0^\dagger+\,a_0\right)
      \,,\qquad
      p=i\sqrt{\pi}(1-\varepsilon)\left(a_0^\dagger-\,a_0\right)\,.
\ee
 They obey a canonical  commutation relation $[\,q\,,\,p\,]=i\,$.
The zero mode part of the BD field is thus expressed as
\be
	\int_{0}^{2\pi}{d\theta \over 2\pi}\,\Phi_{\sst \rm BD}(\eta,\theta)
	=q\,+\,p\,{{\eta-{\pi\over 2}}\over 2\pi}\,+\,{\cal O}(\varepsilon)\,,
\ee
which has a finite limit.

For finite $\varepsilon$ 
the vacuum $|\Omega)$ is annihilated by 
$a_0=\sqrt{\pi}\varepsilon q+i\{2\sqrt{\pi}(1-\varepsilon)\}^{-1}p$. 
When $\varepsilon$ goes to zero
with finite $q$ and $p\,$, this condition reduces to
$p\,|\Omega)=0$ \cite{Allen:1987tz}. Such a state 
cannot be normalized as is the case in the quantum mechanics
for a one-dimensional harmonic oscillator in the limit of vanishing frequency. 
The resulting Fock space is therefore  the tensor product of the 
Fock space we obtained in  Section 4.2  and the  Hilbert space carrying a
representation of  the commutator $[\,q\,,\,p\,]=i\,$.

In the $\varepsilon\to0$ limit 
the zero mode operator $p$ appears also in 
the generators $K_{\pm}$ of the dS group
\be
	K_{+}\,=\,
	{p\over 2\sqrt{\pi}}\left(a_1^\dagger+a_{-1}\right)+\,
	i\sum_{m\neq 0,-1}a^{\dagger}_{m+1}\,a_m\,.
\ee
The condition that the vacuum be dS invariant leads again to $p\,|\Omega)=0$.
It is important to notice that
we no longer are in the general framework we described in Section 3
where the Fock space and the generators of the dS group are constructed only
from the irreducible representations.
The massless limit leads indeed to a larger Fock space
and modified generators: In the subspace where $p=0$, we recover the
generators constructed from the UIR of the discrete series. 
When $p\neq 0$, the number of
particles is no longer dS invariant, i.e. the subspace with $N$ 
particles is no longer
invariant under the dS transformations.

 In the previous section when considering the massless field
from discrete series, we found a unique  field operator  
once we require its  covariance. That  expression  differs from the
above $\Phi_{\sst\rm BD}$. 
This follows from the presence of the zero mode operators
in the generators  of the dS group.
It is therefore of interest to study 
how the zero mode operators transform under the SU$(1,1)$
group. Using eq.(\ref{gener}), the massless limit of the
general field operator is
\ba
	\Phi_{\alpha,\beta}(\eta,\theta)&=&
	q_{\alpha,\beta}\,+\,p_{\alpha,\beta}\,\frac{\eta-\frac\pi2}{2\pi}\,+
	\nn&&+\,
	{1\over 2\sqrt{\pi}}\sum_{m\neq 0}{1\over |m|}
	\left(\cosh\alpha\, e^{i|m|\eta}	+\sinh\alpha\, e^{i\beta}e^{-i|m|\eta}\right)
	e^{-im\theta}a^\dagger_m\,+\,{\rm h.c.}
\ea
where the zero mode operators $q_{\alpha,\beta}$ and
$p_{\alpha,\beta}$ are given by
\ba
	q_{\alpha,\beta}&=&
	{1\over 2\sqrt{\pi}\,\varepsilon}
	\left\{\left(\cosh\alpha+e^{i\beta}\sinh\alpha\right)a_0^\dagger+{\rm h.c}\right\},
	\nonumber\\
	p_{\alpha,\beta}&=&
	i\sqrt{\pi}(1-\varepsilon)
	\left\{\left(\cosh\alpha-e^{i\beta}\sinh\alpha\right)a_0^\dagger-{\rm h.c}\right\},
\ea
The dS generators now read
\be
	K_{+}\,=\,
	{p_{\alpha,\beta}\over 2\sqrt{\pi}}\,
	{a_1^\dagger+a_{-1}\over\cosh\alpha+\cos\beta\sinh\alpha}\,
	+\,i\sum_{m\neq 0,-1}a^{\dagger}_{m+1}\,a_m\,.
\ee
Notice that the generators depend explicitly on $\alpha$ and $\beta$.
Notice also that the dS invariant vacuum must 
satisfy  $p_{\alpha,\beta}\,|\Omega)=0$ which is the same condition as 
$p\,|\Omega)=0$. 
Notice finally that unless $\beta=\pi$,
in the limit $\alpha\rightarrow +\infty$  the zero modes
cancel out from the   generators
and the field operator which for $\beta=0$ reads
\ba
	\Phi_{\alpha,0}(\eta,\theta)&=&q_{\alpha,0}\,+\,
	p_{\alpha,0}\,\frac{\eta- \frac\pi2}{2\pi}\,+
	\nn&&+\,
	{1\over 2\sqrt{\pi}}\sum_{m\neq 0}{1\over |m|}
	\left(i\,e^{\alpha}\sin{|m|\eta}
	+e^{-\alpha}\cos{m\eta}\right) e^{-im\theta}a^\dagger_m
	+{\rm h.c.}
\ea
tends to the non canonical commuting field of eq.(\ref{pi nc field}).

We finally notice that the energy momentum tensor 
given in the light cone coordinates by 
$T_{++}=\ :\partial_+\Phi\,\partial_+\Phi:$
and $T_{--}=\ :\partial_-\Phi\,\partial_-\Phi:$ now
generates two copies of Virasoro algebra:
\be	
	L_m=\int_0^{2\pi} d x^+\ e^{imx^+}\,T_{++}(x^+)\,,
	\qquad
	\tilde L_m=\int_0^{2\pi} d x^-\ e^{imx^-}\,T_{--}(x^-)\,.
\ee
The zero mode generators $L_0$ and $\tilde L_0$ define
the rotation generator:
\be
	J=\tilde L_0-L_0=\sum_{m\ge1}a_{m}^\dagger\,a_{m}-a_{-m}^\dagger\,a_{-m}\,,
\ee
and a Hamiltonian:
\ba
	H=L_0+\tilde L_0
	=\frac{p^2}{4\pi}
	+\sum_{m=1}^\infty &\Bigg[&\cosh2\alpha\left(a_{m}^\dagger\,a_{m}+a_{-m}^\dagger\,a_{-m}\right)+
	\nn&&
	+\sinh2\alpha\left(e^{i\beta}\,a^\dagger_m\,a^\dagger_{-m}+e^{-i\beta}\,a_m\,a_{-m}\right)\Bigg]\,.
\ea
which reproduces the correct time evolution, 
in agreement with the fact that the field is canonical.

\subsection{Vertex operator and massless two-point function}

As we saw in the previous section, the dS invariant vacuum is not normalizable
since it is the solution to $p\,|\Omega)=0$. 
An easy way to regularize
this infinity is to compactify the scalar field on a circle, that is to identify
$\Phi$ and $\Phi+2\pi L$ where $L$ is the radius of the circle. This amounts to
compactify the zero mode $q$ and so
$p$ has discrete eigenvalues $n/L$ with $n\in{\mathbb Z}$ and 
its eigenmodes are normalizable.
Observables should be invariant under $\Phi\rightarrow\Phi+2\pi L\,$.
This excludes the scalar field $\Phi$ as an observable
but includes derivatives of $\Phi$ or a covariant
regularization of $\exp(i{\Phi/L})$ which we will call the vertex operator
$V$.

We define $V$ at the origin of spacetime as the normal ordered
exponential, that is
\be
	V(0)=\,:e^{\frac iL \Phi(0)}:\,,
	\label{vertex}
\ee
where $:\cdot:$ is normal-ordering prescription 
and it does not affect the zero mode.
It can be obtained from the massive fields as
\be
	V(0)=\lim_{\varepsilon\to 0}\,
	e^{\frac iL\Phi^+_\varepsilon(0)}\,
	e^{\frac iL\Phi^-_\varepsilon(0)}\,
	e^{{1\over 4\pi L^2}[\,c_0\,a_0^\dagger\,,\,c_0^*\,a_0\,]}\,.
	\label{vertex lim}
\ee
Here, $\Phi^+_\varepsilon$ (resp. $\Phi^-_\varepsilon$) denotes 
the part of the massive field depending 
on creation (resp. annihilation) operators.
This shows that $V(0,0)$ commutes with $K_2$ because $\Phi^\pm$ do
and the last factor is a $c$-number.
And we can translate with the dS transformations
\be
	V(t,\theta)=e^{-iJ\theta}\,e^{iK_1t}\,V(0,0)\,e^{-iK_1t}\,e^{iJ\theta}\,
	=\lim_{\varepsilon\to 0}
	e^{{i\over L}\Phi^+_\varepsilon(t,\theta)}\,
	e^{{i\over L}\Phi^-_\varepsilon(t,\theta)}\,
	e^{{1\over 4\pi L^2}[\,c_0\,a_0^\dagger\,,\,c_0^*\,a_0\,]}\,,
	\label{vertex}
\ee
and this is an dS covariant definition of the vertex operator.
If we decompose the massless BD field as 
$\Phi(\eta,\theta)=\phi^0(\eta,\theta)+ \phi^+(\eta,\theta)+\phi^-(\eta,\theta)$ with
\be
	\phi^0(\eta,\theta)=q\,+\,p\,{{\eta-{\pi\over 2}}\over 2\pi}\,,
	\quad
	\phi^+(\eta,\theta)={1\over 2\sqrt{\pi}}
	\sum_{m\neq0}\frac{1}{|m|}e^{i|m|\eta-im\theta}\,a^\dagger_m\,,
	\quad
	\phi^-(\eta,\theta)=\left(\phi^+(\eta,\theta)\right)^\dagger\,,
\ee
then the vertex operators defined in eq.(\ref{vertex}) reads
\be
	V(\eta,\theta)\,=\,
	e^{{i\over L}\phi^+(\eta,\theta)}\,
	e^{{i\over L}\phi^-(\eta,\theta)}\,
	e^{{i\over L}\phi^0(\eta,\theta)}\,
	(\sin\eta)^{1\over 4\pi L^2}
	\label{verver}
\ee
It is important to notice that this definition of the vertex operator is not 
same as the usual one $:e^{\frac iL \Phi(x)}:$ 
which was considered in \cite{Casher:2003gc}.
The last factor $(\sin\eta)^{1/4\pi L^2}$ in eq.(\ref{verver})
is necessary for the dS covariance.

The two-point function of vertex operators is readily calculated and is given by 
\be
      (\Omega|\,V^\dagger(x)\,V(x')\ \Omega)=
 	\exp\left[-{1\over 4\pi L^2}\,
	\log\left\{2\left(1-\tilde Z(\bar x;\bar x')\right)\right\}\right]\,,
 \ee
where $\tilde Z(x;x')=
Z(x;x')+i\,{\rm sgn}(t-t')\,\epsilon\,$.
Since $\log z$ has a branch cut on the negative real axis,
the term $i\,\epsilon\,{\rm sgn}(t-t')$ contributes only when $Z(x;x')>1$
and thus is equivalent to a dS invariant quantity
$i\,\epsilon\,s^{(1,2)}(x,x')$ in eq.(\ref{s12}).
Furthermore it  is related to the massive two-point function 
which has a divergence in $(4\pi\varepsilon)^{-1}$ by
\be
	\lim_{\varepsilon\to 0}\;
	\exp\left[\frac1{L^2} \left\{
	(\Omega|\,\Phi_\varepsilon(x)\,\Phi_\varepsilon(x')\,\Omega)-
	\frac1{4\pi}\left(\frac1\varepsilon+2\ln2\right)\right\}\right].
\ee	
Our regularization isolates the divergent piece of 
the two-point function and leaves a dS invariant result.

\section{Arbitrary dimension}

In this section we generalize  to arbitrary dimensions
the approach we have used in two dimensions.
The scalar UIR will be easily generalized 
by their realizations with functions 
on the $(n-1)$-sphere, $S^{n-1}$.

The $n$-dimensional de Sitter space, $dS_n$ is described by the hyperboloid in $(n+1)$-dimensional
Minkowski space ${\mathbb{R}}^{1,n}$:
\be
   \eta_{AB} X^A X^B\,=\,1\,,
   \label{dSn}
\ee
where $\eta_{AB}=diag(-1,1,\dots ,1)$. In the following we use the index notation:
\be
\begin{array}{ll}
       A,B,C,D=0,1,\dots ,n\,; &\quad I,J=1,2,\dots ,n\,;\\
       \mu,\nu=0,1,\dots ,n-1\,;&\quad i,j,k=1,2,\dots ,n-1\,.\\
\end{array}
\ee

\subsection{The complementary series of SO$_0(1,n)$}

The isometry group of $dS_n$ is SO$_0(1,n)$. It
is the group of special orthogonal transformations continuously connected to the identity 
 which leaves eq.(\ref{dSn}) invariant. The generators of SO$_0(1,n)$  verify the following algebra:
\be
       [\,{\cal M}_{AB},\,{\cal M}_{CD}\,]\,=\,-i\,\left(\,
       \eta_{AC}{\cal M}_{BD}-\eta_{AD}{\cal M}_{BC}-\eta_{BC}{\cal M}_{AD}+\eta_{BD}{\cal M}_{AC}
       \,\right).
\ee
The scalar representations of SO$_0(1,n)$ can be realized on the space of 
functions on $S^{n-1}$, which is conveniently parametrized 
by a vector $\vec\zeta$ in $\mathbb{R}^n$ subject to $|\vec\zeta|=1\,$.
The action of the generators of SO$_0(1,n)$ in the representation 
labeled by $s$
 are given by
\ba
       (\,{\scriptstyle \vec\zeta}\,|\,{\cal M}_{IJ}\,\Psi)
	&=&i\left(\zeta_I{\partial \over \partial \zeta^J}-
       \zeta_J{\partial \over \partial \zeta^I}\right)\,(\,{\scriptstyle \vec\zeta}\,|\,\Psi)\,,
	 \nonumber\\
	 (\,{\scriptstyle \vec\zeta}\,|\,{\cal M}_{I0}\,\Psi)
	&=&\left\{\zeta^J  {\cal M}_{IJ}
	 +i\left(s+{n-1\over 2}\right)\zeta_I\right\}\,(\,{\scriptstyle \vec\zeta}\,|\,\Psi)\,,
	\label{n alg}
\ea
and the quadratic Casimir is constant in this representation space and is given by
\be
  	{\cal C}=\frac12 \sum_{A,B} {\cal M}_{AB}\,{\cal M}^{AB}
 	=s^2-\frac{(n-1)^2}4\,,
\ee
with $-(n-1)/2<s<(n-1)/2$ for the complementary series.
Physically, the value
of the Casimir 
is the opposite of the mass squared of the particle $M^2=(n-1)^2/4-s^2\,$.
It will also be useful to have the action under finite transformations, we have
\ba
   &(\,{\st\vec\zeta}\,\vert \,e^{i\theta{\cal M}^{IJ}}\,\Psi)
   &=\,(\,{\st\vec\zeta'}\,\vert \Psi)\,,
   \nonumber\\
   &(\,{\st\vec\zeta}\,\vert \,e^{i\omega{\cal M}^{I0}}\,\Psi)
   &=\,(\cosh\omega+\zeta^I\sinh\omega)^{-s-{n-1\over 2}}
   (\,{\st\vec\zeta''}\,\vert \Psi)\,,
   \label{fdSact}
\ea
where
\ba
	&&\vec\zeta'=e^{i\theta{\mathscr M}^{IJ}}\vec\zeta\,,
	\nonumber\\
	&&\vec\zeta''=\left({\zeta^1\over \cosh\omega+\zeta^I\sinh\omega},\,\dots\,,
   	{\sinh\omega+\zeta^I\cosh\omega\over\cosh\omega+\zeta^I\sinh\omega},\,
   	\dots\,,{\zeta^n\over \cosh\omega+\zeta^I\sinh\omega}\right),
	\label{fdSactw}
\ea
and where ${\mathscr M}^{IJ}$ is the representation of ${\cal M}^{IJ}$ on the vector space ${\mathbb R}^n$.

For purely imaginary $s$, this  representation is unitary with respect to 
the ${\cal L}^2$ scalar product, $(\cdot|\cdot)$, and this is the principal series.
For real $s$, 
the scalar product with respect to which this representation is unitary
is, as in two dimensions, $\la\cdot|\cdot\ra\equiv (\cdot|Q\,\cdot)$
where $Q$ is  the intertwiner operator defined in  eq.(\ref{int}).
From $(\,{\scriptstyle \vec\zeta}\,\vert\,{\cal M}_{AB}^{(-s)}Q\ {\scriptstyle \vec\zeta'}\,)
=(\,{\scriptstyle \vec\zeta'}\,\vert\,{\cal M}_{AB}^{(-s)}Q\ {\scriptstyle \vec\zeta}\,)^*\,$,
the intertwiner is determined up to a normalization:
\be
    	Q({\st\,\vec\zeta}\cdot{\st\vec\zeta'\,})\equiv
	 (\,{\scriptstyle \vec\zeta}\,\vert\,Q\ {\scriptstyle \vec\zeta'}\,)=
	Q_0\,
	\frac{\Gamma\!\left(s+\frac{n-1}2\right)}
      	{(2\pi)^{\frac{n-1}2}\,2^s\,\Gamma(s)}
	\left(1-\vec\zeta\cdot\vec\zeta'\right)^{s-\frac{n-1}2}\,.
	\label{n Q}
\ee
This scalar product is well defined if the value of $s$ is 
restricted to $0<s<(n-1)/2$ and 
this is the complementary series.

The generalization  of the Fourier basis valid in the two dimensional case
is here provided by the spherical harmonics in $n$ dimensions.
These are conveniently defined from the set
of homogeneous polynomials of degree $L$ in
$n$ variables $X_I$ which are harmonic. It can be shown that there are $N(n,L)=
(2L+n-2)(L+n-3)!/L!(n-2)!$ independent harmonic and homogeneous 
polynomials. Their restriction on the unit sphere defines the 
spherical harmonics
$S_{L,k}({\st\vec\zeta})=(\,{\st\vec\zeta}\,|L,k)$ with $L=0,1,\dots,$ and $k=0,1,\dots, N(n,L)\,$.

Notice that ${\cal M}_{IJ}{\cal M}^{IJ}$ is constant on the set of harmonic and homogeneous 
polynomials of degree $L$ and is given by $L(L+n-2)\,$.
It is easy to verify that the operators ${\cal M}_{I\pm}$ defined by 
\be
	(\,{\st\vec\zeta}\,|{\cal M}_{I+}|L,k)=i\left(\zeta_I-\frac{1}{n+2L-2}\partial_I\right)(\,{\st\vec\zeta}\,|L,k)\,,
	\qquad
	 (\,{\st\vec\zeta}\,|{\cal M}_{I-}|L,k)=i\,\partial_I(\,{\st\vec\zeta}\,|L,k)\,,
	\label{ladder n}
\ee
change the degree of $L$ by $\pm1$. Then the boost
can be deduced from the expression of the generators as
\be
	{\cal M}_{I0}\,|L,k)=\left\{\left(L+s+{n-1\over 2}\right){\cal M}_{I+}
	+\frac{s-L-{n-3\over2}}{L+n-1}\,{\cal M}_{I-}\right\}|L,k)\,,
	\label{boo} 
\ee
thus the boost generators ${\cal M}_{I0}$ change the degree $L$ of $|L,k)$ by $\pm 1$
while the rotation generators ${\cal M}_{IJ}$ leave the degree $L$ unchanged.
Since the intertwiner $Q$ commutes with the rotations, it is
of the form $Q\,|L,k)=Q_L\,|L,k)$ where $Q_L$ are $c$-numbers determined from
the commutation with ${\cal M}_{I0}\,$. Using eq.(\ref{boo}), we get
\be
	Q_L=\la L_0|L_0 \ra\, {\Gamma(s+{n-1\over 2}+L_0)\,\Gamma(-s+{n-1\over 2}+L)\over
	\Gamma(-s+{n-1\over 2}+L_0)\,\Gamma(s+{n-1\over 2}+L)}\,.\label{ql}
\ee
Here we introduced a reference state $L_0$ with a norm $\la L_0|L_0 \ra\,$.
In the following we shall take $L_0=1$ and $\la\,1\,|\,1\,\ra=1\,$, this also
fixes the
  normalization constant
$Q_0$ in eq.(\ref{n Q})  by $Q_0=(s+\frac{n-1}{2})/(-s+\frac{n-1}{2})\,$.
Notice that the expression (\ref{ql}) is valid for $s$ positive and negative
in the interval $\,\mathop{]\!-\!(n-1)/2,\,(n-1)/2[}$ whereas the expression (\ref{n Q}) applies
only for positive $s$.
The above expression of the intertwiner in spherical harmonic basis
can be also obtained also from eq.\eqref{n Q} as is shown in Appendix A.

\subsection{The massless representation of SO$_0(1,n)$}

It is the first member of the discrete series with a vanishing
quadratic Casimir operator.
It can be realized on functions on the sphere
$S^{n-1}$ with vanishing zero modes. A basis is thus given
by the spherical harmonics with $L\ge1\,$.
The action of the generators is given by eq.(\ref{n alg}) with
$s=(n-1)/2\,$. The rotations do not change the degree of homogeneity
and the action of the boosts is given by
\be
	{\cal M}_{I0}\,|L,k)=\left\{(L+n-1)\,{\cal M}_{I+}
	+\frac{L-1}{L+n-1}\,{\cal M}_{I-}\right\}|L,k)\,,
	\label{boom} 
\ee
where ${\cal M}_{I\pm}$ were defined in eq.(\ref{ladder n}).
For $L=1$, the right-hand side is of degree $L=2$,
the set of harmonics with strictly positive degree is thus invariant by the
action of the generators.
The generators will be Hermitian with respect to  
the scalar product $(\cdot|Q\,\cdot)$, with $Q$ deduced as for the complementary
series with $s=(n-1)/2$:
\be
	Q_L={(n-1)!(L-1)!\over (L+n-2)!}\,.
\ee
Notice that the limit $s\to(n-1)/2$ of the complementary series
decomposes into an invariant state ($L=0$) and the massless representation.
The limit is singular because the norm of the zero mode varies as
$\Gamma(s-{\st n-1\over 2})$ which becomes infinite.

As noticed before, an equivalent representation is provided by $s=-(n-1)/2\,$.
The scalar product is now given by
\be
	Q_L={(L+n-2)!\over (n-1)!(L-1)!}\,.
\ee
 The limit $s\to -(n-1)/2$ of the  norm of the zero mode
 of the complementary series is now zero making this description of the
 massless representation more convenient.
However in this case, the action of the boosts on the $L=1$ state generates
the zero norm $L=0$ state:
\be
	{\cal M}_{I0}\,|1,k)=\left({\cal M}_{I+}
	-\frac{n-1}{n}\,{\cal M}_{I-}\right)|1,k)\,.
\ee

\subsection{Scalar field from the complementary series}

We choose the origin in $dS_n$ to have the embedding coordinates $X_o^A=(0,0,\dots ,0,1)$.
The field on any point of $dS_n$, of coordinates  $X^A=\Lambda^A_{\phantom{A}B} X_o^B$,
 can be deduced from the field at the origin $\Phi(X_o)$ by
\be
   \Phi(X)=U(\Lambda) \, \Phi(X_o) \, U(\Lambda)^{-1}\,,
   \label{fondeq}
\ee
where $\Lambda$ is an element of SO$_0(1,n)$.

In the global coordinate system:
\ba
   X^0&=&\sinh t\,,\nonumber\\
   X^I&=&\cosh t\,\xi^I\,,\qquad\vec\xi\in S^{n-1}\,,\nonumber\\
   \xi^I &=&{\mathrm{R}}^I_{\phantom{I}J}\,\xi_o^J\,,\qquad\vec\xi_o=(0,\dots,0,1)\,,
\ea
where $\mathrm{R}$ is a element of SO$(n)$ subgroup. The metric reads
\be
   ds^2=-dt^2+\cosh^2 t\, d\Omega^2({\st\,\vec\xi\,})\,.
\ee
Therefore the point $X_o$ can be transported to any point $X$ by a boost followed by a rotation.
This implies that eq.(\ref{fondeq}) can be written as
\be
   \Phi(X)=U({\mathrm{R}})\, e^{itM^{0n}}\,\Phi(X_o)\,e^{-itM^{0n}}\,U({\mathrm{R}})^{-1}\,.
\ee
The origin $X_o=(0,\vec\xi_o)$ is invariant under the action of the subgroup
SO$_0(1,n-1)$ generated by $M_{\mu\nu}$. To construct a local field, we thus require
\be
   [\,M_{\mu\nu},\,\Phi(0,{\st\vec\xi_o})\,]=0\,.
   \label{cvn}
\ee

From the UIR of SO$_0(1,n)$ we define the creation and annihilation operators by
\ba
       &a^\dagger({\scriptstyle\,\vec\zeta\,})\,\vert\Omega)
	=\vert{\scriptstyle\,\vec\zeta\,})\,,
       \qquad
       a({\scriptstyle\,\vec\zeta\,})\,\vert\Omega)=0\,,
        \\
       &[\,a({\scriptstyle\,\vec\zeta\,}),\, a^\dagger({\scriptstyle\,\vec\zeta'\,})\,]
	=Q({\scriptstyle\,\vec\zeta}\cdot{\scriptstyle\vec\zeta'\,})\,,
       \qquad\,
	[\,a({\scriptstyle\,\vec\zeta\,}),\, a({\scriptstyle\,\vec\zeta'\,}) \,]=0\,.
\ea
The field operator in the origin can be expanded in terms of these operators:
\be
   \Phi(0,{\st\vec\xi_o})=\int d^{n-1}\Omega({\st\,\vec\zeta\,})\
  \Psi_o({\st\,\vec \zeta\,})\,a^\dagger({\st\,\vec\zeta\,})\,+\,\Psi_o^*({\st\,\vec\zeta\,})\,a({\st\,\vec\zeta\,})\,,
\ee
where $d^{n-1}\Omega({\st\,\vec\zeta\,})$ is the invariant volume element on $S^{n-1}$.

The covariance condition (\ref{cvn}) determines
 the function $\Psi_o({\st\,\vec\zeta\,})=({\scriptstyle\,\vec\zeta\,}\,|\,\Phi(0,{\scriptstyle\,\vec\xi_o\,})\,\Omega)$. 
The rotation part of this equation implies
 that $\Psi_o({\st\,\vec\zeta\,})$ depends on $\vec\zeta$ only through $\vec\zeta\cdot\vec\xi_o$,
 whereas the boost part fixes this dependence to be  again governed by two arbitrary coefficients:
\be
     \Psi_{o}({\scriptstyle\,\vec\zeta\,})=
	C\,\left(\vec\zeta\cdot\vec\xi-i\epsilon\right)^{-s-\frac{n-1}2}
     +D\,\left(\vec\zeta\cdot\vec\xi+i\epsilon\right)^{-s-\frac{n-1}2}\,.
\ee
Transporting the field with a boost followed by a rotation
brings the point $(0,{\st\vec\xi_o})$ to
$(t,{\st\vec\xi\,})$. We thus get
\be
       \Phi(t,{\scriptstyle\,\vec\xi\,})\,=\,\int_{S^{n-1}} d^{n-1}\Omega({\scriptstyle\,\vec\zeta\,})\
	\Psi_{t,{\scriptscriptstyle\,\vec\xi\,}}({\scriptstyle\,\vec\zeta\,})\,a^\dagger({\scriptstyle\,\vec\zeta\,})
        \,+\, \Psi_{t,{\scriptscriptstyle\,\vec\xi\,}}^*({\scriptstyle\,\vec\zeta\,})\,a({\scriptstyle\,\vec\zeta\,})\,,
\ee
where $\Psi_{t,{\scriptscriptstyle\,\vec\xi\,}}({\scriptstyle\,\vec\zeta\,})
=({\scriptstyle\,\vec\zeta\,}\,|\,\Phi(t,{\scriptstyle\,\vec\xi\,})\,\Omega)$ is given by
\be
     \Psi_{t,{\scriptscriptstyle\,\vec\xi\,}}({\scriptstyle\,\vec\zeta\,})=
	C\,\left(\cosh t\,\vec\zeta\cdot\vec\xi+\sinh t-i\epsilon\right)^{-s-\frac{n-1}2}
     +D\,\left(\cosh t\,\vec\zeta\cdot\vec\xi+\sinh t+i\epsilon\right)^{-s-\frac{n-1}2}\,.
	\label{psi n}
\ee
At this point we are comparing with the results of the principal series I.
The expression (\ref{psi n}) is identical with that of the principal series
with $s\to -i\mu$. The only difference is to be found in the commutation relations 
of creation and annihilation operators.

The two-point function is simply expressed in terms of
$\Psi_{t,\xi}({\st\,\vec\zeta\,})$ and the intertwiner 
$Q({\scriptstyle\,\vec\zeta\,\cdot\,\vec\zeta'\,})$ as
\be
       	(\Omega\vert\,\Phi(t,{\scriptstyle\,\vec\xi\,})\,\Phi(t',{\scriptstyle\,\vec\xi\,}')\,\Omega)=
	\int_{S^{n\!-\!1}}\!\!\int_{S^{n\!-\!1}}\!\!\!d^{n\!-\!1}\Omega({\scriptstyle\,\vec\zeta\,})\,
	d^{n\!-\!1}\Omega({\scriptstyle\,\vec\zeta\,}')\ 
       \Psi^*_{t,{\scriptscriptstyle\,\vec\xi\,}}({\scriptstyle\,\vec\zeta\,})\,
	Q({\scriptstyle\,\vec\zeta\,\cdot\,\vec\zeta'\,})\,
	\Psi_{t',{\scriptscriptstyle\,\vec\xi\,}'}({\scriptstyle\,\vec\zeta\,}')\,.
	\label{n 2pt fn}
\ee
 The above integral can be expressed in terms of hypergeometric functions as shown
 in the Appendix B. We obtain
\be
       (\Omega|\,\Phi(x)\,\Phi(x')\,\Omega)=|C|^2\,F_n(x;x')+|D|^2\,F_n(\bar x;\bar x') +
	2\,{\rm Re}\!\left[C^*D\,e^{i\pi\left(-s-\frac{n-1}2\right)} F_n(x;\bar x')\right],
	\label{twop} 
\ee
 with
\be
	 F_n(x;x')=V_{n-1}\,\frac{s+\frac{n-1}{2}}{-s+\frac{n-1}{2}}\, 
	{}_2F_1\left(s+\frac{n-1}2\,,\,-s+\frac{n-1}2\,;\,\frac n2\,;
	\,\frac{1+\tilde Z(x;x')}2\right),
\ee
where $V_{n-1}$ is the area of $(n-1)$-sphere, $2\pi^{n/2}/\Gamma(n/2)\,$.
We have defined $\tilde Z$ by
\be
	\tilde Z(x;x')=Z(x;x')+i\,{\rm sgn}(t-t')\,\epsilon\,,
	\label{tildeZ}
\ee
where $Z$ is the dS invariant quantity:
\be
	Z(t,\vec\xi;t',\vec\xi')=
	\cosh t\cosh t'\,\vec\xi\cdot\vec\xi'-
   	\sinh t\sinh t'=X^AX'_A\,.
\ee

The analysis of the equal time commutation relations made in I 
for the principal series applies also here and leads to commuting fields and 
\be
   \left[\,\partial_t \Phi(0,{\st\vec\xi_o\,}),\,\Phi(0,{\st\vec\xi\,})\,\right]=iN_n\,\delta^{n-1}({\st\,\vec\xi-\vec\xi_o\,})\,,
\ee
with $N_n$ a constant which we shall shortly determine in terms of $C$ and $D$.
From the imaginary part of the short distance behavior of two-point function (\ref{twop}) which is given by
\ba
 	G_{dS_n}(x;x')&\underset{ x\to x'}{\approx}&
 	\frac{2^{n-1}\,\pi^{\frac n2}\left(s+\frac{n-1}{2}\right)}
	{\Gamma\left(s+\frac{n-1}2\right)\Gamma\left(-s+\frac{n+1}2\right)}
	\times
	\nonumber\\
	&&\times
 	\left\{
 	\begin{array}{ll}
    	|C|^2\,\log\left\{(x-x')^2-(t-t'+i\epsilon)^2\right\}^{-2}\,+&\\
	\quad+\,|D|^2\,\log\left\{(x-x')^2-(t-t'-i\epsilon)^2\right\}^{-2}\,,&\;{\rm for}\;n=2\,,\\
    	&\\
    	|C|^2\left\{(\vec x-\vec x')^2-(t-t'+i\epsilon)^2\right\}^\frac{2-n}2+&\\
    	\quad+\,|D|^2\left\{(\vec x-\vec x')^2-(t-t'-i\epsilon)^2\right\}^\frac{2-n}2\,,&\;{\rm for}\;n\neq2\,,
 	\end{array} \right.
\ea
the requirement $N_n=1$ leads to
\be
|D|^2-|C|^2=
	\frac{\Gamma\left(s+\frac{n-1}2\right)\Gamma\left(-s+\frac{n+1}2\right)}
	{2^{n+1}\pi^{n}\left(s+\frac{n-1}{2}\right)}\,.
	 \label{ccrn}
\ee

Another constraint on $C$ and $D$ can be obtained by demanding that the two
point function coincides with the flat one in the small distance limit.
This gives the so called Bunch-Davies vacuum \cite{BD}.
 In the coincidence point limit,
the Minkowski vacuum positive Wightman function behaves as
\be
 G_{\rm{Mink}}(x;x')\underset{ x\to x'}{\approx}
 \frac1{4\pi^{n\over 2}}\times
 \left\{
 \begin{array}{ll}
    \log\left\{(x-x')^2-(t-t'-i\epsilon)^2\right\}^{-2}\,, &\;{\rm for}\; n=2\,,\\
    &\\
    \Gamma({n\over 2}-1)
    \left\{(\vec x-\vec x')^2-(t-t'-i\epsilon)^2\right\}^\frac{2-n}2\,, &\;{\rm for}\; n\neq2\,.
 \end{array} \right.
\ee
 Imposing that the behavior 
 be that of Hadamard, one gets the Bunch-Davies coefficients:
\be
 	C^{\mathrm{BD}}=0\,,
 	\qquad D^{\mathrm{BD}}=
 	\sqrt{\frac{\Gamma\left(s+\frac{n-1}2\right)\Gamma\left(-s+\frac{n+1}2\right)}
	{2^{n+1}\pi^{n}\left(s+\frac{n-1}{2}\right)}}\,.
	\label{nbd}
\ee
The general coefficients $C$ and $D$ can be expressed in terms of the Bunch-Davies coefficients as
\be
       C=e^{i\beta}\sinh\alpha\,e^{-i\pi(s+\frac{n-1}{2})}D^{\rm{BD}}\,,\qquad
       D=\cosh\alpha\, D^{\rm{BD}}\,.
	\label{n alpha}
\ee

To study the asymptotic time behavior of the field operator
and determine the $\alpha$ and $\beta$ coefficients corresponding to the \emph{in} and
\emph{out} vacua, it is useful to expand
the field  in the spherical harmonic basis as
\be
	\Phi(t,{\scriptstyle\,\vec\xi\,})\,=\,
	\sum_{L,k}\,c_L(t)\,S_{L,k}({\scriptstyle\,\vec\xi\,})\,a^\dagger_{L,k}+{\rm h.c.}
\ee
where the creation operators $a^\dagger_{L,k}$ are related to UIR states $|L,k)\,$.
The coefficient $c_L(t)$ is calculated 
from eq.(\ref{psi n}) by decomposing on the 
spherical harmonics as  shown in Appendix A, it reads
\ba
	c_{L}(t)=
	\frac{2\left(\frac{2\pi}{\cosh t}\right)^{\frac{n-1}{2}}}{\Gamma(s+\frac{n-1}{2})}&\Bigg[&
	D\,e^{-i\pi(s+\frac{n-1}{2})}\,
	\left({\rm Q}^{s}_{\sst L+\frac{n-3}{2}}(-\tanh t)
	+i\,\frac{\pi}{2}\,{\rm P}^{s}_{\sst L+\frac{n-3}{2}}(-\tanh t)\right)+
	\nn
	&&
	+\,C\,e^{i\pi(s+\frac{n-1}{2})}\,
	\left({\rm Q}^{s}_{\sst L+\frac{n-3}{2}}(-\tanh t)
	-i\,\frac{\pi}{2}\,{\rm P}^{s}_{\sst L+\frac{n-3}{2}}(-\tanh t)\right)
	\Bigg]\,,
	\nn
	\label{cL}
\ea
where ${\rm P}^{\mu}_{\nu}$ and ${\rm Q}^{\mu}_{\nu}$ are 
associated Legendre functions of the first and second kind respectively.
Using the parametrization (\ref{n alpha}),
the asymptotic behavior 
in the remote past of $c_L(t)$ for non-integer $s$
 is shown in the Appendix A to be given by
\be
	c_L(t)\underset{t\to-\infty}=P(e^{2t})\,e^{\left(s+\frac{n-1}2\right)t}
	\left\{1+i\,\omega^{\sst\rm in}_L\,
	\left(2\,e^{-2st}\right)+{o}\left(e^{-2s t}\right)\right\},
	\label{asym cL}
\ee	
where $P$ is a real polynomial  of degree 
less than $-s$ up to an overall phase
and the frequency $\omega^{\sst\rm in}_L$ is given by
\be
	\omega^{\sst\rm in}_L=
	\frac{\Gamma(1+s)\,\Gamma(L-s+\frac{n-1}{2})}
	{2\,\Gamma(1-s)\,\Gamma(L+s+\frac{n-1}{2})}
	\frac{-\sin\pi s +i(\cos\pi s \cosh 2\alpha+\cos(\beta+\frac{n-1}{2}\pi)\sinh 2\alpha)}
	{\cosh2\alpha+\cos\left(\beta+\left(s+\frac{n-1}{2}\right)\pi\right)\sinh2\alpha}\,.
	\label{w in}
\ee
Factorizing the overall decreasing term and defining 
a new time coordinate as eq.(\ref{redef time}),
the last factor of  $c_L(t)$ 
can be written as a plane wave to the first order in $\eta_s$ 
as the two-dimensional case (\ref{in plane}).
The \emph{in} vacuum with respect to $\eta_s$ is defined 
by $\omega^{\sst\rm in}_L$ real and 
positive. Since the real part of $\omega^{\sst\rm in}_L$ is always positive,
the condition   ${\rm Im}[\omega^{\sst\rm in}_L]=0$ is 
sufficient
and reads
\be
	\cos\pi s=-\tanh2\alpha\,\cos\left(\beta+\frac{n-1}{2}\pi\right)\,.
\ee
This condition 
interpolates between the BD vacuum in the conformally massless case
which in $n$-dimensions corresponds to $s=-1/2$
and the Mottola-Schwinger \emph{in} vacuum of the principal 
series which is given with the conditions \cite{Mottola:1984ar}:
\be
	\cosh\pi \mu=\coth 2\alpha\,,
	\qquad
	\cos\left(\beta+\frac{n-1}{2}\pi\right)=-1\,,
\ee
where $\mu=i\,s\in\mathbb{R}\,$.

The large future limit is similarly determined in the Appendix to be 
given by
\be
	c_L(t)\underset{t\to\infty}=Q(e^{-2t})\,
	e^{-\left(s+\frac{n-1}2\right)t}
	\left\{1+i\,\omega^{\sst\rm out}_L\,
	\left(-2\,e^{2st}\right)+{o}\left(e^{2s t}\right)\right\},
\ee	
where $Q$ is a real polynomial of degree less than $-s$ multiplied by an overall
constant and where $\omega^{\sst\rm out}_L$ is given by
\be
	\omega^{\sst\rm out}_L=
	\frac{\Gamma(1+s)\,\Gamma(L-s+\frac{n-1}{2})}
	{2\,\Gamma(1-s)\,\Gamma(L+s+\frac{n-1}{2})}
	\frac{-\sin\pi s +i(\cos\pi s \cosh 2\alpha+\cos(\beta-\frac{n-1}{2}\pi)\sinh 2\alpha)}
	{\cosh2\alpha+\cos\left(\beta-\left(s+\frac{n-1}{2}\right)\pi\right)\sinh2\alpha}\,.
\ee
Using a time coordinate $\eta_s$ given in eq.(\ref{redef time out}),
the last factor of $c_L(t)$ becomes a plane wave to the first order in $\eta_s$
and the reality of  $\omega^{\sst\rm out}_L$ 
defines the \emph{out} vacuum:
\be
	\cos\pi s=-\tanh2\alpha\,\cos\left(\beta-\frac{n-1}{2}\pi\right)\,.
\ee
 The resulting  $\omega^{\sst\rm out}_L$ is positive.
 Comparing with the \emph{out} vacuum in the principal series:
\be
	\cosh\pi \mu=\coth 2\alpha\,,
	\qquad
	\cos\left(\beta-\frac{n-1}{2}\pi\right)=-1\,,
\ee
we can see that this newly defined \emph{out} vacuum also interpolates
between the BD vacuum in the conformally massless case $s=-1/2$
and the Mottola-Schwinger \emph{out} vacuum
in the principal series.

Notice that contrary to the case of the principal series 
we get a family of dS invariant \emph{in} and \emph{out} vacua parametrized by $\beta$.
For the time-reversal \emph{in} and \emph{out} vacua with
$\alpha^{\sst\rm out}=\alpha^{\sst\rm in}$, $\beta^{\sst\rm out}=-\beta^{\sst\rm in}$,
the mean number 
of \emph{out} quanta of momentum $L$ in \emph{in} vacuum 
can be easily obtained as
\be
	\bar{n}_{\sst\rm out/in}
	=\cosh^2 (2\alpha^{\sst\rm in})\,\cos^2\pi s
	=\frac{\cos^2\left(\beta^{\sst\rm in}-\frac{n-1}{2}\pi\right)\,\cos^2\pi s}
	{\cos^2\left(\beta^{\sst\rm in}-\frac{n-1}{2}\pi\right)-\cos^2\pi s}\,.
\ee
It vanishes, as it should, in the conformal case $s=-1/2$ 
and diverges in the $s\to 0$ limit.
Notice that the number also vanishes for $s$ half integer and tends to infinity
when $s$ approaches an integer.
It should also be noticed that for $n$ odd the family of \emph{in} vacua
coincides with the family of \emph{out} vacua, this is to be compared with the 
fact that in odd dimensions the \emph{in} and \emph{out}
 vacua also coincide for the principal series. 
Notice that when $\beta^{\rm in}=(n-1)\pi/2\,$, the number of created quanta is minimal
and is given by $\cot^ 2 \pi s$. It is easy to show that this is
also the minimal number of created quanta when considering
all \emph{in} and  \emph{out} vacua.

It remains to examine the case of integer $s$, in this case
the asymptotic behavior of $c_L(t)$ in the remote past is given by
\be
	c_L(t)\underset{t\to-\infty}=R(e^{2t})\,e^{\left(s+\frac{n-1}2\right)t}
	\left\{1+i\,\nu_L\,
	\left(2\,e^{-2st}\right)+{o}\left(e^{-2s t}\right)\right\},
	\label{asym int}
\ee	
where $R$ is a decreasing real function with overall complex constant 
and the frequency $\nu_L$ is given by
\ba
	\nu_L&=&
	\frac{\Gamma(L+\frac{n-1}{2}-s)}{2\,\Gamma(-s)\,\Gamma(1-s)\,\Gamma(L+\frac{n-1}{2}+s)}
	\Bigg[\pi\,
	\frac{1-i\sinh2\alpha\sin(\beta+s+\frac{n-1}{2})}{\cosh2\alpha+\cos(\beta+s+\frac{n-1}{2})\sinh2\alpha}+
	\nn&&
	-i\Big\{\psi(1)+\psi(1-s)-\psi(L+{\st\frac{n-1}{2}})-\psi(L+{\st\frac{n-1}{2}}-s)\Big\}\Bigg]\,,
\ea
where $\psi$ is digamma function.
Requiring $\nu_L$ to be real, we obtain a constraint on the value of $\alpha$ and $\beta$
as in the case of non-integer $s$. However in the present case, 
the condition that $\alpha$ and $\beta$ should satisfy depends also on the mode $L\,$,
which means thus defined \emph{in} vacuum is not dS invariant.

\subsection{Massless limit of massive field}

We consider the massless limit of the massive scalar field from 
the complementary series.
As $\varepsilon=s+(n-1)/2$ approaches to zero,
the zero mode coefficient, $c_0(t)$, in eq.(\ref{cL})
diverges, since the associated Legendre function of the second kind 
${\rm Q}^{\varepsilon-\frac{n-1}{2}}_{\sst L+\frac{n-3}{2}}$
diverges when $\varepsilon$ approaches zero.

This divergence can be also more explicitly seen in the position basis.
If we expand $\Psi_{\eta, \sst \vec\xi}$ given in eq.(\ref{psi n}) 
in $\varepsilon$, using
eq.(\ref{nbd}) and eq.(\ref{n alpha}), we get 
\ba
	&&\Psi_{\eta, \sst \vec\xi}({\st\,\vec\zeta\,})=\sqrt{\frac{\Gamma(n)}{2^{n+1}\pi^{n}}}
	\times\nn
	&&\quad\times
	\Bigg[\left(\cosh\alpha+e^{i\beta}\sinh\alpha\right)
	\left(\frac1\varepsilon+\ln(\sin\eta)-\ln\left|\vec\zeta\cdot\vec\xi-\cos\eta\right|
	+\frac12(\psi(1)-\psi(n))-i\frac{\pi}{2}\right)+
	\nonumber\\
	&&\qquad\qquad
	+\,i\,\pi\left(\cosh\alpha-e^{i\beta}\sinh\alpha\right)
	\left\{\frac{1}{2}-\Theta\left(\cos\eta-{\st\vec\zeta\cdot\vec\xi\,}\right)\right\}\Bigg]
	\,+\,{\cal O}(\varepsilon)\,.
	\label{zee}
\ea
The only divergent term in the $\varepsilon\to0$ limit is a constant, 
i.e. it contributes 
only to the zero mode $c_0(\eta)$.
Using eq.(\ref{zee}), we get 
the small $\varepsilon$ behavior of the zero mode  as
\ba
	c_0(\eta)=\sqrt{\frac{V_{n-1}}{V_{n}}}&\Big[&
	\left(\cosh\alpha+e^{i\beta}\sinh\alpha\right)\left(\frac1\varepsilon+f(\eta)-i\frac{\pi}{2}\right)
	\nn
	&&
	+\,i\left(\cosh\alpha-e^{i\beta}\sinh\alpha\right)g(\eta)\,\Big]
	+{\cal O}(\varepsilon)\,,
	\label{0mode}
\ea
where $f$ and $g$ are defined by
\ba
	f(\eta)&=&\ln(\sin\eta)+\frac12\left(\psi(1)-\psi(n)\right)
	-\frac{V_{n-2}}{V_{n-1}}\int_0^\pi d\phi\,\sin^{n-2}\phi\,\ln|\cos\eta-\cos\phi|\,,
	\nn
	g(\eta)&=&\frac{\pi\,V_{n-2}}{V_{n-1}}\int^\eta_{\frac{\pi}{2}} d\phi\,\sin^{n-2}\phi
	=-\frac{\pi\,V_{n-2}}{V_{n-1}}\cos\eta\;
	{}_2F_1\left(\frac{1}{2}\,,\,\frac{3-n}{2}\,;\,\frac{1}{2}\,;\,\cos^2\eta\right).
\ea 
Another source of singularity in the $\varepsilon\to 0$ limit
is the vanishing of the norm of the zero mode. This translates into the
vanishing of the commutator:
\be
	[\,a_0\,,\,a^\dagger_0\,]=\frac\varepsilon{n-1-\varepsilon}\,.
\ee
The combination $(c_0\,a_0^\dagger +c^*_0\,a_0)/\sqrt{V_{n-1}}$ which appears
in the zero mode part of the field operator can be put in the form
 \be
 q_{\sst \alpha, \beta}^{}\,+\,p_{\sst \alpha, \beta}^{}\,\frac{g(\eta)}
 {\pi\,V_{n-2}}
 \ee
 where $q_{\sst \alpha, \beta}$ and $p_{\sst \alpha, \beta}$ are given by
\ba
	q_{\sst \alpha, \beta}^{}&=&
	\frac{1}{\varepsilon\sqrt{V_{n}}}
	\left\{ \left(\cosh\alpha+e^{i\beta}\sinh\alpha\right)  a_0^\dagger+{\rm h.c.}\right\},
	\nn
	p_{\sst \alpha, \beta}^{}&=&i\sqrt{V_{n}}\,\left(\frac{n-1-\varepsilon}{2}\right)
	\left\{ \left(\cosh\alpha-e^{i\beta}\sinh\alpha\right)   a_0^\dagger-{\rm h.c.}\right\}\,,
\ea
and they obey canonical commutation relations
 $[\,q_{\sst \alpha, \beta}^{}\,,\,p_{\sst \alpha, \beta}^{}\,]=i\,$.
Finally, the scalar field in the massless limit is  expressed as
\be
	\Phi(\eta,{\st\vec\xi})
	\,=\,q_{\sst \alpha, \beta}^{}\,+\,p_{\sst \alpha, \beta}^{}\,\frac{g(\eta)}{\pi\,V_{n-2}}\,
	+\int_{S^{n-1}} d^{n-1}\Omega({\st\vec\zeta})\,
	\Upsilon_{\eta,\vec\xi}({\st\vec\zeta})\,a_*^\dagger({\st\vec\zeta})
	+{\rm h.c.}
\ee
where the non-zero part of creation operator $a_*^\dagger({\st\vec\zeta})$ and its 
coefficient $\Upsilon_{\eta,\vec\xi}$ are defined by
\ba
	&&
	{a}_*^\dagger({\st\vec\zeta})\equiv
	a^\dagger({\st\vec\zeta})-\frac1{\sqrt{V_{n-1}}}\,a^\dagger_0\,,
	\qquad
	[\,a_*({\st\,\vec\zeta\,})\,,\,a_*^\dagger({\st\,\vec\zeta\,}')\,]
	=Q({\st\vec\zeta}\cdot{\st\vec\zeta'})\,,
	\nn
	&&
	\Upsilon_{t, \sst \vec\xi}({\st\vec\zeta})=-\sqrt{\frac{\Gamma(n)}{2^{n+1}\pi^{n}}}\;
	\Big[\,e^{i\beta}\sinh\alpha\,\ln\left(\cosh t\vec\zeta\cdot\vec\xi-\sinh t-i\epsilon\right)+
	\nn
	&&\hskip 4 cm
	+\cosh\alpha\, \ln\left(\cosh t\vec\zeta\cdot\vec\xi-\sinh t+i\epsilon\right)\Big]\,.
\ea
The above expression has a well defined limit.

As in the two-dimensional case, as $\varepsilon$ goes to zero
with finite $q_{\sst \alpha, \beta}^{}$ and $p_{\sst \alpha, \beta}^{}\,$, 
the defining condition of the zero mode 
vacuum state reduces to
$p_{\sst \alpha, \beta}^{}\,|\Omega)=0$. Such a state 
cannot be normalizable and 
the resulting Fock space is therefore  the tensor product of the 
Fock space constructed from massless representation 
and the Hilbert space carrying a
representation of  
$[\,q_{\sst \alpha, \beta}^{}\,,\,p_{\sst \alpha, \beta}^{}\,]=i\,$, i.e. the
one describing a one dimensional quantum mechanical  particle.

The resulting generators of the dS group
are deformed by $p_{\sst \alpha, \beta}$:
in the complementary series 
the zero mode part of $M_{I0}$ can be  written as
\be
	M_{I0}^{(0)}=\sum_{J=1}^n\,(1,J|\,M_{I0}\,|\,0)\,\frac{1}{Q_0}\,a^\dagger_{1,J}\,a_0\,+\,
	(0\,|\,M_{I0}\,|1,J)\,\frac{1}{Q_1}\,a^\dagger_0\,a_{1,J}\,,
	\label{n boost}
\ee
where $|\,0)$ and $|1,J)$ are the states in the spherical harmonic basis:
\be
	({\st\,\vec\zeta\,}|\,0)=S_0({\st\,\vec\zeta\,})=\frac{1}{\sqrt{V_{n-1}}}\,,
	\qquad
	({\st\,\vec\zeta\,}|\,1,J)=S_{1,J}({\st\,\vec\zeta\,})=\sqrt{\frac{n}{V_{n-1}}}\,\zeta^J\,,
\ee
In terms of the redefined zero mode operators $p_{\sst \alpha, \beta}^{}$ and $q_{\sst \alpha, \beta}^{}$, eq.(\ref{n boost})
has a limit given by
\be
	M_{I0}^{(0)}=\frac{1}{\sqrt{n\,V_{n}}}\,
	\frac{p_{\alpha,\beta}^{}\left(a^\dagger_{1,I}+a_{1,I}\right)}{\cosh \alpha+\cos\beta\sinh\alpha}\,.
\ee
This part should be added to the generators obtained from the massless UIR of
the dS group.
The dS invariance of the vacuum state leads 
again $p_{\sst \alpha, \beta}^{}\,|\Omega)=0\,$, making the vacuum state
non-normalizable.

\subsection{Vertex operator}

In order to make the  vacuum state  normalizable, 
we compactify the scalar field
on a circle of radius $L\,$ as in the two-dimensional case. 
The physical  observables should then be invariant 
under $\Phi \to\Phi+2\pi L$. This is the case for the vertex operator
$V$ which is a dS covariant regularization of $\exp(i\Phi/L)\,$.

As in two dimensional case, this regularization can be 
realized by defining $V$ at the origin as the normal ordered exponential
\eqref{vertex} and transporting with the dS transformations.
The  resulting vertex operator can be expressed as  
\ba
	V(\eta,{\st\vec\xi})&=&
	\exp\left(\frac{i}{L}\phi^+(\eta,{\st\vec\xi})\right)
	\exp\left(\frac{i}{L}\phi^-(\eta,{\st\vec\xi})\right)
	\exp\left(\frac{i}{L}\phi^0(\eta,{\st\vec\xi})\right)
	\times\nn
	&&\times
	\exp\left(\frac1{2 V_{n-1} L^2}\lim_{\varepsilon\to0}\left(|c_0(\eta)|^2-|c_0({\st\frac\pi2})|^2\right)
	[\,a_0\,,\,a_0^\dagger\,]\right),
	\label{vern}
\ea
where $\phi^+$, $\phi^-$ and $\phi^0$ are respectively 
creation, annihilation and zero mode part of scalar field operator.
Notice that it differs from the normal ordered exponential which is not
dS invariant. The difference is a time dependent constant given in the 
last factor in  eq.(\ref{vern}) and it reads explicitly
\be
	\exp\left[\frac{	(\cosh2\alpha+\sinh2\alpha\cos\beta)\left(f(\eta)-f({\st\frac\pi2})\right)
	+\sinh2\alpha\sin\beta\left(g(\eta)-g({\st\frac\pi2})\right)}{2\pi\,V_{n-2}\,L^2}\right].
\ee

Two-point function of the vertex operators 
$(\Omega|\,V^\dagger(x)\,V(x')\,\Omega)\equiv
\exp\left(\frac1{L^2}\,{\cal G}_n(x;x')\right)$
can now be easily calculated 
and expressed in terms of the limit of the massive two-point function as
\be
	{\cal G}_n(x;x')
	=\lim_{\varepsilon\to0}\left\{
	 (\Omega|\,\Phi_\varepsilon(x)\,\Phi_\varepsilon(x')\,\Omega)-
	\frac{|c_0(t=0)|^2}{V_{n-1}}[\,a_0\,,\,a_0^\dagger\,]\right\}.
\ee
Using the two-point function
(\ref{twop}) and  the zero mode (\ref{0mode}), the divergent 
part cancels and we get a finite limit 
\ba
	2\pi\,V_{n-2}\,{\cal G}_n(x;x')&=&
	\sinh^2\alpha\, G_n(x;x')+\cosh^2\alpha\, G_n(\bar x;\bar x')
	\nn	
	&&+\,2\,{\rm Re}
	\left[\,\sinh\alpha\cosh\alpha\,e^{-i\beta}\,G_n(x;\bar x')\,\right],
	\label{2pfv}
\ea
where the dS invariant function $G_{n}(x;x')$ is defined by
\be
	G_{n}(x;x')=
	\left.
	\frac{\partial}{\partial\varepsilon}\,
	{}_2F_1\left(-\varepsilon+n-1\,,\,\varepsilon\,;\,\frac n2\,;
	\,\frac{1+\tilde Z(x;x')}2\right)
	\right|_{\varepsilon=0}
	+\psi\left(\frac{n-1}{2}\right)-\psi\left(\frac{n}{2}\right).
	\label{Gn}
\ee
Notice that the procedure we propose results 
in the extraction from the massive two
point function of its divergent part. In fact,
$(\Omega|\,\Phi_\varepsilon(x)\,\Phi_\varepsilon(x')\,\Omega)$
 behaves in the massless limit as
\be
	\frac{\cosh2\alpha+\sinh2\alpha\cos\beta}{2\pi\,V_{n-2}}\,
	\left(\frac{1}{\varepsilon}+\psi(1)-\psi(n-1)
	-\psi\left(\frac{n-1}{2}\right)+\psi\left(\frac{n}{2}\right)\right)
	+{\cal G}_n(x;x')\,.
\ee
 The function $G_n(x;x')=f(Z(x;x'))$ can be determined 
from the differential equation 
satisfied by ${}_2F_1(-\varepsilon+n-1\,,\,\varepsilon\,;\,n/2\,;
\,(1+\tilde Z(x;x'))/2)$ which results in
\be
	(1-x^2)\,f''(x)-n\,x\,f'(x)-(n-1)=0\,,\label{eqg}
\ee
and its explicit expression is given in Appendix C.

An interesting and unexpected property of ${\cal G}_n$ is its behavior for large
$|Z|$,
that is in the far infrared region in the flat sections.
In this limit eq.(\ref{eqg}) simplifies to
$(x^nf')'=-(n-1)\,x^{n-2}$ and its solution for large $x$ is
$-\log x$ independently of $n$. 
The \emph{massless two-point function} has thus a logarithmic divergence 
for all spacetime dimensions similar to the two dimensional flat infrared
divergence. 
This IR divergence was found in \cite{Ratra:1984yq} by considering the BD massless
two-point function and disregarding its divergent part.
Here, this regularization arises naturally in a dS invariant way and is due to
the compactification of the scalar on a circle.
In this respect, this IR divergence was argued to cause
a restoration of a breakdown of symmetry \cite{Ratra:1984yq} similarly to
what happens in two dimensions (For discussions on IR divergence for the graviton,
see e.g. \cite{ATH}).
The two-point function
of the vertex operators can be used to exhibit this symmetry restoration.
From our previous analysis  for $x$ and $x'$ separated by a large spacelike 
distance $d(x;x')=\cosh ^{-1}Z(x;x')$, we have
\be
	(\Omega|\,V^\dagger(x)\,V(x')\,\Omega) \sim 
	Z(x;x')^{-\frac{\cosh2\alpha+\sinh2\alpha\cos\beta}{2\pi\,V_{n-2}\,L^2}},
\ee
which tends to zero. This is a signal of large quantum fluctuations
which restore a broken symmetry.

\appendix

\section{Intertwiner and Field coefficients in spherical harmonic basis}

In this subsection, we derive the intertwiner $Q_L$ and the field coefficient $c_L(t)$
in the spherical harmonic basis.
Many properties of special functions used in the following
can be found in \cite{bateman,Abramowitz}

At first, we concentrate on the case of the intertwiner operator.
From the $n$-dimensional addition theorem  which reads
\be
	\sum_{k=1}^{N(n,L)}(\,{\st\vec\xi}\,|L,k)(L,k|\,{\st\vec\zeta}\,)=
	\sum_{k=1}^{N(n,L)}S_{L,k}({\st\vec\xi})S_{L,k}({\st\vec\zeta})=
	\frac{2L+n-2}{(n-2)V_{n-1}}\,C_{L}^{n-2\over2}({\st\vec\zeta\cdot\vec\xi})\,,
	\label{add th}
\ee
with $C_{L}^{n-2\over2}$ the Gegenbauer polynomial, 
we get the relation between the intertwiners in the position basis and 
in the spherical harmonic basis:
\be
	Q({\st\vec\zeta\cdot\vec\zeta'})=
	\sum_{L,k}Q_L\,S_{L,k}({\st\vec\zeta})\,S_{L,k}({\st\vec\zeta'})
	=\sum_{L}Q_L\,\frac{2L+n-2}{(n-2)V_{n-1}}\,C_{L}^{n-2\over2}({\st\vec\zeta\cdot\vec\zeta'})\,.
	\label{itgr Q}
\ee
Using the orthogonality of the Gegenbauer polynomials, we express $Q_L$ as
\be
	Q_L=(4\pi)^{\frac{n-2}{2}}\frac{\Gamma(L+1)\,\Gamma(\frac{n-2}{2})}{\Gamma(L+n-2)}\,\int_{-1}^{1}dx
	\, (1-x^2)^{\frac{n-3}{2}}C_{L}^{\frac{n-2}{2}}(x)\,Q(x)\,.
\ee
On the other hand, the Gegenbauer polynomials are given by
 \be
	C_{L}^{\frac{n-2}{2}}(x)=\frac{(-1)^L\,\pi^{\frac{1}{2}}\,2^{-L-n+3}\,\Gamma(L+n-2)}
	{\Gamma(L+\frac{n-1}{2})\,\Gamma(L+1)\,\Gamma(\frac{n-2}{2})}
	(1-x^2)^{-\frac{n-3}{2}}\left(\frac{d}{dx}\right)^L(1-x^2)^{L+\frac{n-3}{2}}\,.
	\label{Gegen}
\ee
Using the expression of $Q({\st\vec\zeta\cdot\vec\zeta'})$ in eq.(\ref{n Q})
and applying an integration by part, we get
\be
	Q_L=Q_0\,\frac{\Gamma(s+\frac{n-1}{2})\,\Gamma(-s+\frac{n-1}{2}+L)}
	{2^{s+\frac{n-3}{2}+L}\Gamma(s)\,\Gamma(L+\frac{n-1}{2})\,\Gamma(-s+\frac{n-1}{2})}
	\int_{-1}^{1}dx\, (1-x)^{s-\frac{n-1}{2}-L}(1-x^2)^{L+\frac{n-3}{2}}\,.
\ee
Performing the integral we get an expression of $Q_L$ which coincides eq.(\ref{ql}).

The coefficient of field operator $c_L(t)$ can also be obtained in a similar manner.
We first decompose $\Psi_{t,{\sst\,\vec\xi\,}}({\sst\,\vec\zeta\,})$ in the Gegenbauer polynomials as
\ba
	\Psi_{t,{\sst\vec\zeta}}({\st\vec\xi})&=&
	\sum_{L,k}({\scriptstyle\,\vec\zeta\,}\,|L,k)(L,k|\,\Phi(t,{\scriptstyle\,\vec\xi\,})\,\Omega)
	=\sum_{L,k}c_L(t)\,S_{L,k}({\st\vec\zeta})\,S_{L,k}({\st\vec\xi})
	\nn
	&=&
	\sum_{L}c_L(t)\,\frac{2L+n-2}{(n-2)V_{n-1}}\,C_{L}^{n-2\over2}({\st\vec\zeta\cdot\vec\xi})\,,
\ea
and using their orthogonality, we express $c_L(t)$ as an integral:
\be
	c_L(t)=(4\pi)^{\frac{n-2}{2}}\frac{\Gamma(L+1)\,\Gamma(\frac{n-2}{2})}{\Gamma(L+n-2)}\,\int_{-1}^{1}dx
	\, (1-x^2)^{\frac{n-3}{2}}C_{L}^{\frac{n-2}{2}}(x)\,\Psi(t,x)\,,
\ee
where $\Psi(t,x)=C\,(\cosh t\,x+\sinh t-i\epsilon)^{-s-\frac{n-1}2}+D\,(\cosh t\,x+\sinh t+i\epsilon)^{-s-\frac{n-1}2}\,$.
This integral can be again evaluated easily using eq.(\ref{Gegen}) and 
the integral representation of the hypergeometric function. Finally we get
\ba
	c_{L}(t)&=&\frac{\pi^{\frac{n+2}{2}}}{\Gamma\left(L+\frac{n}{2}\right)}
	\left(\frac{\cosh t}{2}\right)^{L}e^{\left(s+\frac{n-1}{2}+L\right)t}\times
	\\
	&&\times
	\Bigg[
	D\,e^{-i\pi\left(s+\frac{n-1}{2}\right)}
	{}_2F_1\left(L+\frac{n-1}{2}\,,\,L+\frac{n-1}{2}+s\,;\,2L+n-1\,;\,1+e^{2t}+i\epsilon\right)+
	\nn
	&&
	\qquad+\,C\,e^{i\pi\left(s+\frac{n-1}{2}\right)}
	{}_2F_1\left(L+\frac{n-1}{2}\,,\,L+\frac{n-1}{2}+s\,;\,2L+n-1\,;\,1+e^{2t}-i\epsilon\right)\Bigg]\,.
	\nonumber
	\label{cLt}
\ea
Since the hypergeometric functions ${}_2F_1(a,b;c;z)$ have a branch cut on ${\rm Arg}[z-1]=0\,$, 
the above two hypergeometric functions with different $i\epsilon$ prescriptions are independent. 
Using the transformation identities between the hypergeometric and the associated Legendre functions, 
the above expression of  $c_{L}(t)$ can be written also as eq.(\ref{cL}).

In the subsection 6.3, we used the asymptotic behavior of $c_L(t)$
in order to define \emph{in} and \emph{out} vacua.
These asymptotic behaviors can be obtained from 
those of the hypergeometric function.
We concentrate on the large past case, $t\to-\infty$ and the 
large future case can be done in a similar manner.
When $s$ is not a integer and in the interval $\,\mathop{]\!-\!m\!-\!1,-m[}\,$, 
the asymptotic behavior of the hypergeometric function is given by
\ba
	&&{}_2F_1\left(L+\frac{n-1}{2}\,,\,L+\frac{n-1}{2}+s\,;\,2L+n-1\,;\,1+e^{2t}-i\epsilon\right)\nn
	&&=\frac{\Gamma(2L+n-1)\,\Gamma(-s)}{\Gamma(L+\frac{n-1}{2})\,\Gamma(L+\frac{n-1}{2}-s)}
	\Bigg(\sum_{l=0}^{m}\frac{(L+\frac{n_1}{2})_{(l)}(L+\frac{n-1}{2}+s)_{(l)}}{(1+s)_{(l)}\,l!}(-1)^{l}e^{2lt}+\nn
	&&\qquad 
	+\,\frac{\Gamma(s)\,\Gamma(L+\frac{n-1}{2}-s)}{\Gamma(-s)\,\Gamma(L+\frac{n-1}{2}+s)}e^{-i\pi s}e^{-2st}
	\,+\,{o}(e^{-2st})\,\Bigg)\nn
	&&=
	p(e^{2t})\,\left(
	1+\frac{\Gamma(s)\,\Gamma(L+\frac{n-1}{2}-s)}
	{\Gamma(-s)\,\Gamma(L+\frac{n-1}{2}+s)}e^{-i\pi s}e^{-2st}
	+{o}(e^{-2st})\right)\,,
	\label{asym F}
\ea
where $p(e^{2t})$ is a real polynomial of order $m$ in $e^{2t}$ and given by
\be
	p(e^{2t})=\frac{\Gamma(2L+n-1)\,\Gamma(-s)}{\Gamma(L+\frac{n-1}{2})\,\Gamma(L+\frac{n-1}{2}-s)}
	\sum_{l=0}^{m}\frac{(L+\frac{n-1}{2})_{(l)}(L+\frac{n-1}{2}+s)_{(l)}}{(1+s)_{(l)}\,l!}(-1)^{l}e^{2lt}\,.
\ee
Combining the two hypergeometric functions with coefficients $C$ and $D$, we get
the asymptotic behavior of $c_L(t)$ as
\ba
	&&c_L(t)=\frac{\pi^{\frac{n+2}{2}}}{4^L\Gamma\left(L+\frac{n}{2}\right)}
	\left(D\,e^{-i\pi\left(s+\frac{n-1}{2}\right)}+C\,e^{i\pi\left(s+\frac{n-1}{2}\right)}\right)
	p(e^{2t})\,e^{\left(s+\frac{n-1}{2}\right)t}
	\times\nn
	&&\times\left(1+\frac{D\,e^{-i\pi\frac{n-1}{2}}+C\,e^{i\pi\frac{n-1}{2}}}
	{D\,e^{-i\pi\left(s+\frac{n-1}{2}\right)}+C\,e^{i\pi\left(s+\frac{n-1}{2}\right)}}
	\frac{\Gamma(s)\,\Gamma(L+\frac{n-1}{2}-s)}
	{\Gamma(-s)\,\Gamma(L+\frac{n-1}{2}+s)}e^{-i\pi s}e^{-2st}+{o}(e^{-2st})\right),\nn
\ea
this gives eq.(\ref{asym cL}) after using eq.(\ref{w in}) and 
defining $P(e^{2t})$ as $p(e^{2t})$ times the overall constant of $c_L(t)$ in eq.(\ref{cLt}).

For the case of integer $s$, the expression (\ref{asym F}) is replaced by
\ba
	&&\frac{\Gamma(2L+n-1)}{\Gamma(L+\frac{n-1}{2})}
	\Bigg[\frac{\Gamma(-s)}{\Gamma(L+\frac{n-1}{2}-s)}
	\sum_{l=0}^{-s-1}\frac{(L+\frac{n-1}{2})_{(l)}(L+\frac{n-1}{2}+s)_{(l)}}{(1+s)_{(l)}\,l!}(-1)^{l}e^{2lt}
	\nn&&
	-\frac{e^{-2st}}{\Gamma(L+\frac{n-1}{2}+s)\,\Gamma(1-s)}
	\Big\{2t+i\pi-\psi(1)-\psi(1-s)+\psi(L+{\st\frac{n-1}{2}})+\psi(L+{\st\frac{n-1}{2}}-s)\Big\}\Bigg]
	\nn&&
	=r(e^{2t})\,\Bigg(1-\frac{\Gamma(L+\frac{n-1}{2}-s)}{\Gamma(L+\frac{n-1}{2}+s)}
	\frac{e^{-2st}}{\Gamma(-s)\,\Gamma(1-s)}\times
	\nn&&\hskip 100pt \times
	\Big\{i\pi-\psi(1)-\psi(1-s)+\psi(L+{\st\frac{n-1}{2}})+\psi(L+{\st\frac{n-1}{2}}-s)\Big\}\Bigg),
\ea
where $r(e^{2t})$ is given by
\ba
	&&r(e^{2t})=\frac{\Gamma(2L+n-1)}{\Gamma(L+\frac{n-1}{2})}
	\frac{\Gamma(-s)}{\Gamma(L+\frac{n-1}{2}-s)}\times\\
	&&\times\left(
	\sum_{l=0}^{-s-1}\frac{(L+\frac{n-1}{2})_{(l)}(L+\frac{n-1}{2}+s)_{(l)}}{(1+s)_{(l)}\,l!}(-1)^{l}e^{2lt}
	-\frac{2\,\Gamma(L+\frac{n-1}{2}-s)}{\Gamma(L+\frac{n-1}{2}+s)\,\Gamma(-s)\,\Gamma(1-s)}\,e^{-2st}\,t
	\right).\nonumber
\ea
Combining again two hypergeometric functions with $C$ and $D$,
the asymptotic behavior of $c_L(t)$ with integer value of $s$ is given by eq.(\ref{asym int}),
where $R(e^{2t})$ is $r(e^{2t})$ times the overall constant in eq.(\ref{cLt}).

\section{Wightman function}

The two-point function in $n$-dimensions given in eq.(\ref{n 2pt fn}) reads
\ba
	&&(\Omega|\,\Phi(t,{\st\,\vec\xi\,})\,\Phi(t',{\st\,\vec\xi\,}')\,\Omega)=
	Q_0\,\frac{\Gamma\left(s+\frac{n-1}2\right)}{(2\pi)^{\frac{n-1}2}\,2^s\,\Gamma(s)}\,\times
	\nonumber\\
	&&\qquad \times
	\int_{\sst S^{n\!-\!1}\times S^{n\!-\!1}}\!\!\!\!\!\!
	d^{n\!-\!1}\Omega({\st\,\vec\zeta\,})\,d^{n\!-\!1}\Omega({\st\,\vec\zeta\,}')\ 
 	\left(1-\vec\zeta\cdot\vec\zeta'\right)^{s-\frac{n-1}2}
	\Psi^*_{t,{\sst\,\vec\xi\,}}({\st\,\vec\zeta\,})\,\Psi_{t',{\sst\,\vec\xi\,}'}({\st\,\vec\zeta\,}')\,,
\ea
with $\Psi_{t,{\sst\,\vec\xi\,}}({\st\,\vec\zeta})$ given before.
Each term with coefficients $|C|^2$, $|D|^2$, $C^*D$ and $D^*C$ can be written in terms of a
 single function $F_n$ with appropriate arguments as
\ba
	&&(\Omega|\,\Phi(t,{\st\,\vec\xi\,})\,\Phi(t',{\st\,\vec\xi\,}')\,\Omega)
	=|C|^2\,F_n(x;x')+|D|^2\,F_n(\bar x;\bar x')+
	\nonumber\\
	&&\qquad\qquad
	+\,C^*D\,e^{i\pi\left(-s-\frac{n-1}2\right)}\,F_n(x;\bar x')
	\,+\,D^*C\,e^{-i\pi\left(-s-\frac{n-1}2\right)}\,F_n(\bar x;x')\,,
\ea
with
\ba
	&&F_n(t,{\st\,\vec\xi\,};t',{\st\,\vec\xi\,}')=Q_0\,
	\frac{\Gamma\left(s+\frac{n-1}2\right)}{2^s\,(2\pi)^{\frac{n-1}2}\,\Gamma(s)}\times
	\nn&&\qquad\quad\times
	\int_{S^{n-1}}\int_{S^{n-1}} d^{n-1}\Omega(\vec\zeta)\,d^{n-1}\Omega(\vec\zeta')\,
	\left(1-\vec\zeta\cdot\vec\zeta'\right)^{s-\frac{n-1}2}\times
	\nonumber\\
	&&\qquad\quad\times
	\left(\cosh t\,\vec\zeta\cdot\vec\xi+\sinh t+i\epsilon\right)^{-s-\frac{n-1}2}\,
	\left(\cosh t'\,\vec\zeta'\cdot\vec\xi'+\sinh t'-i\epsilon\right)^{-s-\frac{n-1}2}\,.
\ea
Using the following integral obtained in I:
\ba
	&&\int d^{n-1}\Omega(\vec{\zeta})\,
	\left(\cosh t\,\vec\zeta\cdot\vec\xi+\sinh t+i\epsilon\right)^{s-\frac{n-1}2}\,
	\left(\cosh t'\,\vec\zeta\cdot\vec\xi'+\sinh t'-i\epsilon\right)^{-s-\frac{n-1}2}
	\nonumber\\
	&&\qquad=\,e^{i\pi s}\frac{2\pi^{\frac n2}}{\Gamma(\frac n2)}\,
	{}_2F_1\left(i\mu+\frac{n-1}2,-i\mu+\frac{n-1}2;\frac n2;
	\frac{1+\tilde Z(t,\vec\xi;t',\vec\xi')}2\right)\,,
\ea
where
\be
     \tilde Z(t,\vec\xi;t',\vec\xi')=\cosh t\cosh t'\,\vec\xi\cdot\vec\xi'-
     \sinh t\sinh t'+i\,\mathrm{sgn}(t-t')\,\epsilon\,,
\ee
the integral with respect to $\vec\zeta'$ can be expressed as a limit of hypergeometric function:
\ba
	&&\int_{S^{n-1}}d^{n-1}\Omega(\vec\zeta')\,
	\left(1-\vec\zeta\cdot\vec\zeta'\right)^{s-\frac{n-1}2}
	\left(\cosh t'\,\vec\zeta'\cdot\vec\xi'+\sinh t'-i\epsilon\right)^{-s-\frac{n-1}2}
	\nn&&
	=(e^{-i\pi})^{s-\frac{n-1}2}
	\lim_{t\to-\infty}\Bigg\{ (2e^{t})^{s-\frac{n-1}2}
	\int_{S^{n-1}}d^{n-1}\Omega(\vec\zeta')\,
	\left(\cosh t\,\vec\zeta\cdot\vec\zeta'+\sinh t+i\epsilon\right)^{s-\frac{n-1}2}
	\times\nn&&
	\hskip7cm\times
	\left(\cosh t'\,\vec\zeta'\cdot\vec\xi'+\sinh t'-i\epsilon\right)^{-s-\frac{n-1}2}
	\Bigg\}
	\nonumber\\
	&&=\,(e^{-i\pi})^{s-\frac{n-1}2}\lim_{t\to-\infty}\Bigg\{(2e^{t})^{s-\frac{n-1}2}
	e^{i\pi s}\frac{2\pi^{\frac n2}}{\Gamma(\frac n2)}
	\times\nn&&
	\hskip4cm\times
	{}_2F_1\left(s+\frac{n-1}2,-s+\frac{n-1}2;\frac n2;
	\frac{1+\tilde Z(t,\vec\zeta;t',\xi')}2\right)\Bigg\}.
\ea
From the asymptotic behavior of hypergeometric function:
\ba
	&&{}_2F_1\left(s+\frac{n-1}2,-s+\frac{n-1}2;\frac n2;\frac{1+x}2\right)
	\nonumber\\
	&&\quad\underset{|x|\to\infty}{\approx}
	\frac{\Gamma\left(\frac n2\right)\Gamma(2s)}
	{\Gamma\left(s+\frac{n-1}2\right)\Gamma\left(s+\frac12\right)}
	\left(-\frac x2\right)^{s-\frac{n-1}2}
	+\frac{\Gamma\left(\frac n2\right)\Gamma(-2s)}
	{\Gamma\left(-s+\frac{n-1}2\right)\Gamma\left(-s+\frac12\right)}
	\left(-\frac x2\right)^{-s-\frac{n-1}2}\,.
	\nonumber\\
\ea
we get the integral as
\be
	e^{i\pi s}\frac{2^s\,(2\pi)^{\frac{n-1}2}\,\Gamma(s)}{\Gamma\left(s+\frac{n-1}2\right)}
	\,\left(\cosh t'\,\vec\zeta\cdot\vec\xi'+\sinh t'-i\epsilon\right)^{s-\frac{n-1}2}\,,
\ee
Finally, the integral with respect to $\vec\zeta$ gives
\ba
	F_n(t,\vec\xi;t',\vec\xi')&=&Q_0\,e^{i\pi s}
	\int_{S^{n-1}} d^{n-1}\Omega(\vec\zeta)
	\left(\cosh t\,\vec\zeta\cdot\vec\xi+\sinh t+i\epsilon\right)^{-s-\frac{n-1}2}\times
	\nn&&\hskip4cm\times
	\left(\cosh t'\,\vec\zeta\cdot\vec\xi'+\sinh t'-i\epsilon\right)^{s-\frac{n-1}2}\,
	\nonumber\\
	&=&Q_0\,V_{n-1}\,
	{}_2F_1\left(s+\frac{n-1}2,-s+\frac{n-1}2;\frac n2;\frac{1+\tilde Z}2\right)\,.
\ea

\section{Explicit expression of  the two-point function for vertex operators}

The two-point function of vertex operators, 
$(\Omega|\,V^\dagger(x)\,V(x')\,\Omega)\equiv
\exp\left(\frac1{L^2}\,{\cal G}_n(x;x')\right)$
was given by eq.(\ref{2pfv}) and eq.(\ref{Gn}).
An explicit expression for $G_n$ can be deduced from the differential equation (\ref{eqg})
and the boundary conditions at $Z =\pm1$ 
which gives for $n$ even:
\ba
	G_{n}(x;x')&=&
	-\log\left\{2\left(1-\tilde{Z}(x;x')\right)\right\}+
	\sum_{m=1}^{\frac{n-2}{2}}
	\frac{\left(\frac{2-n}{2}\right)_m}{m\,(2-n)_m}
	\left\{\frac{2^m}{\left(1-\tilde{Z}(x;x')\right)^m}-1\right\}+
	\nn
	&&\ +\,\psi\left(\frac{n-1}{2}\right)-\psi\left(\frac{n}{2}\right)+
	\psi\left(1\right)-\psi\left(\frac{1}{2}\right),
\ea
and for $n$ odd:
\ba
	G_{n}(x;x')&=&\frac{n-1}n\,\left(1-\tilde{Z}(x;x')\right)
	{}_3F_2\left(1\,,\,1\,,\,n\,;\,2\,,\,1+\frac n2\,;\,\frac{1-\tilde{Z}(x;x')}2\right)+
	\\
	&&+\,\frac{\Gamma\left(\frac n2\right)\Gamma\left(\frac n2\right)}
	{2^{n+1}\,\Gamma(n-1)}\,\tilde{Z}(x;x')\,
	{}_2F_1\left(\frac12\,,\,\frac n2\,;\,\frac32\,;\,\left(\tilde{Z}(x;x')\right)^2\right)
	+\psi\left(\frac{n-1}{2}\right)-\psi\left(\frac{n}{2}\right).
	\nonumber
\ea
These two expressions have a logarithmic behavior in the IR at large $|Z|\,$.

\acknowledgments

We are grateful to X. Bekaert, E. Buffenoir, J.P. Gazeau, K. Noui, and
 Ph. Roche for stimulating discussions.

\bibliographystyle{JHEP.bst}

\bibliography{ref}

\providecommand{\href}[2]{#2}\begingroup\raggedright\begin{thebibliography}{10}

\bibitem{Wigner:1939cj}
E.~P. Wigner, {\it On unitary representations of the inhomogeneous lorentz
  group},  {\em Annals Math.} {\bf 40} (1939) 149--204.

\bibitem{Weinberg:1995mt}
S.~Weinberg, {\it The quantum theory of fields. vol. 1: Foundations}, .
  Cambridge, UK: Univ. Pr. (1995) 609 p.

\bibitem{Weinberg:1996kw}
S.~Weinberg, {\it What is quantum field theory, and what did we think it was?},
   \href{http://xxx.lanl.gov/abs/hep-th/9702027}{{\tt hep-th/9702027}}.

\bibitem{Joung:2006gj}
E.~Joung, J.~Mourad, and R.~Parentani, {\it Group theoretical approach to
  quantum fields in de sitter space. i: The principal series},  {\em JHEP} {\bf
  08} (2006) 082, [\href{http://xxx.lanl.gov/abs/hep-th/0606119}{{\tt
  hep-th/0606119}}].

\bibitem{Bargmann:1946me}
V.~Bargmann, {\it Irreducible unitary representations of the lorentz group},
  {\em Annals Math.} {\bf 48} (1947) 568--640.

\bibitem{gel}
M.~N. IM~Gelfand, {\it Unitary representations of the proper lorentz group},
  {\em Izv. Akad. Nauk SSSR} {\bf 11} (1947) 411.

\bibitem{th}
L.~Thomas, {\it On unitary representations of the group of the de sitter
  group},  {\em Annals Math.} {\bf 42} (1941) 113.

\bibitem{nt}
T.~Newton, {\it A note on the representation of the de sitter group},  {\em
  Annals Math.} {\bf 52} (1950) 730.

\bibitem{rep2}
J.~Dixmier, {\it Representations integrables du groupe de de sitter},  {\em
  Bull. Soc. Math. France} {\bf 89} (1961) 9.

\bibitem{rep3}
A.~W. Knapp and E.~M. Stein, {\it Interwining operators for semisimple groups},
   {\em Ann. of Math.(2)} {\bf 93} (1971) 489.

\bibitem{rep4}
T.~O. Philipps and E.~P. Wigner, {\it De sitter space and positive energy}, .
  in: Group theory and its applications, ed. by E.\ M.\ Loebl, New York and
  London Academic Press (1968) 631.

\bibitem{rep5}
J.~Mickelsson and J.~Niederle, {\it Contractions of representations of the de
  sitter groups},  {\em Commun. Math. Phys.} {\bf 27} (1972) 167.

\bibitem{rep6}
E.~Thieleker, {\it The unitary representations of the generalized lorentz
  groups},  {\em Transactions of the American mathematical Society} {\bf 199}
  (1974) 327.

\bibitem{Allen:1985ux}
B.~Allen, {\it Vacuum states in de sitter space},  {\em Phys. Rev.} {\bf D32}
  (1985) 3136.

\bibitem{Mottola:1984ar}
E.~Mottola, {\it Particle creation in de sitter space},  {\em Phys. Rev.} {\bf
  D31} (1985) 754.

\bibitem{Folacci:1996dv}
A.~Folacci, {\it Toy model for the zero mode problem in the conformal sector of
  de sitter quantum gravity},  {\em Phys. Rev.} {\bf D53} (1996) 3108--3117.

\bibitem{Ford:1977in}
L.~H. Ford and L.~Parker, {\it Infrared divergences in a class of
  robertson-walker universes},  {\em Phys. Rev.} {\bf D16} (1977) 245--250.

\bibitem{Allen:1987tz}
B.~Allen and A.~Folacci, {\it The massless minimally coupled scalar field in de
  sitter space},  {\em Phys. Rev.} {\bf D35} (1987) 3771.

\bibitem{Polarski:1991ek}
D.~Polarski, {\it Infrared divergences in de sitter space},  {\em Phys. Rev.}
  {\bf D43} (1991) 1892--1895.

\bibitem{Kirsten:1993ug}
K.~Kirsten and J.~Garriga, {\it Massless minimally coupled fields in de sitter
  space: O(4) symmetric states versus de sitter invariant vacuum},  {\em Phys.
  Rev.} {\bf D48} (1993) 567--577,
  [\href{http://xxx.lanl.gov/abs/gr-qc/9305013}{{\tt gr-qc/9305013}}].

\bibitem{Folacci:1992xc}
A.~Folacci, {\it Zero modes, euclideanization, and quantization},  {\em Phys.
  Rev.} {\bf D46} (1992) 2553--2559.

\bibitem{Gazeau:1999mi}
J.~P. Gazeau, J.~Renaud, and M.~V. Takook, {\it Gupta-bleuler quantization for
  minimally coupled scalar fields in de sitter space},  {\em Class. Quant.
  Grav.} {\bf 17} (2000) 1415--1434,
  [\href{http://xxx.lanl.gov/abs/gr-qc/9904023}{{\tt gr-qc/9904023}}].

\bibitem{Unruh:1975gz}
W.~G. Unruh, {\it Particle detectors and black holes}, . In *Trieste 1975,
  Proceedings, Marcel Grossmann Meeting On General Relativity*, Oxford 1977,
  527-536.

\bibitem{Ratra:1984yq}
B.~Ratra, {\it Restoration of spontaneously broken continuous symmetries in de
  sitter space-time},  {\em Phys. Rev.} {\bf D31} (1985) 1931--1955.

\bibitem{BD}
T.~S. Bunch and P.~C.~W. Davies, {\it Quantum field theory in de sitter space:
  Renormalization by point splitting},  {\em Proc. Roy. Soc. Lond.} {\bf A360}
  (1978) 117--134.
\\
N.~D. Birrell and P.~C.~W. Davies, {\it Quantum fields in curved space}, .
  Cambridge, Uk: Univ. Pr. (1982) 340p.

\bibitem{Green:1987sp}
M.~B. Green, J.~H. Schwarz, and E.~Witten, {\it Superstring theory. vol. 1:
  Introduction}, . Cambridge, Uk: Univ. Pr. (1987) 469 P. ( Cambridge
  Monographs On Mathematical Physics).

\bibitem{Casher:2003gc}
A.~Casher, P.~O. Mazur, and A.~J. Staruszkiewicz, {\it De sitter invariant
  vacuum states, vertex operators, and conformal field theory correlators},
  \href{http://xxx.lanl.gov/abs/hep-th/0301023}{{\tt hep-th/0301023}}.

\bibitem{ATH}
I.~Antoniadis and E.~Mottola, {\it Graviton fluctuations in de sitter space},
  {\em J. Math. Phys.} {\bf 32} (1991) 1037--1044.
\\
N.~C. Tsamis and R.~P. Woodard, {\it Strong infrared effects in quantum
  gravity},  {\em Ann. Phys.} {\bf 238} (1995) 1--82.
\\
A.~Higuchi and R.~H. Weeks, {\it The physical graviton two-point function in de
  sitter spacetime with s(3) spatial sections},  {\em Class. Quant. Grav.} {\bf
  20} (2003) 3005--3022, [\href{http://xxx.lanl.gov/abs/gr-qc/0212031}{{\tt
  gr-qc/0212031}}].

\bibitem{bateman}
{\it Bateman manuscript project, higher transcendental functions vol. i.}, . A.
  Erd\'elyi, Editor, McGraw-Hill Book Co, Inc.\ New York (1953).

\bibitem{Abramowitz}
M.~Abramowitz and I.~A. Stegun, {\it Handbook of mathematical functions}, .
  National Bureau of Standard Applied Mathematics Series 55 (19).

\end{thebibliography}\endgroup

\end{document}